\documentclass[journal=jacsat,manuscript=article]{achemso}

\SectionsOn         
\SectionNumbersOn

\usepackage{xcolor}               
\usepackage{soul}                  
\sethlcolor{yellow}                
\usepackage{etoolbox}             
\robustify\ref
\robustify\cite

\soulregister\ref  7               
\soulregister\cite 7
\soulregister\eqref 7
\soulregister\ce   7               

\usepackage{xcolor}          
\usepackage{soul}            
\sethlcolor{yellow}

\usepackage[version=3]{mhchem}  

\usepackage{hyperref}

\author{Yu Wang}
\email{18yu.wang@tum.de}
\affiliation{Technical University of Munich, CIT, Department of Computer Science, Boltzmannstra{\ss}e 3, 85748 Garching, Germany}

\author{Maxine Luo}
\email{man.luo@mpq.mpg.de}
\affiliation{Max Planck Institute of Quantum Optics, Hans-Kopfermann-Stra{\ss}e 1, 85748 Garching, Germany}
\affiliation{Munich Center for Quantum Science and Technology, Schellingstra{\ss}e 4, 80799 M\"unchen, Germany}

\author{Matthias Reumann}
\email{matthias.reumann@tum.de}
\affiliation{Technical University of Munich, CIT, Department of Computer Science, Boltzmannstra{\ss}e 3, 85748 Garching, Germany}

\author{Christian B.~Mendl}
\email{christian.mendl@tum.de}
\affiliation{Technical University of Munich, CIT, Department of Computer Science, Boltzmannstra{\ss}e 3, 85748 Garching, Germany}
\affiliation{Technical University of Munich, Institute for Advanced Study, Lichtenbergstra{\ss}e 2a, 85748 Garching, Germany}

\title[An \textsf{achemso} demo]
  {Enhanced Krylov Methods for Molecular Hamiltonians: Reduced Memory Cost and Complexity Scaling via Tensor Hypercontraction}

\usepackage[english]{babel}

\usepackage[ascii]{inputenc}
\usepackage{amsmath,amssymb,amsfonts,amsthm}
\usepackage{mathtools}
\usepackage{braket}
\usepackage{graphicx}
\usepackage{float}
\usepackage{tikz-network}
\usepackage{tikz}
\usepackage{caption}
\usepackage{subcaption}
\usepackage{algorithm}
\usepackage{algpseudocode}
\usepackage{float}

\algrenewcommand\algorithmicdo{}

\usetikzlibrary{decorations.pathreplacing, decorations.text}

\DeclareUnicodeCharacter{00F6}{\"o}

\DeclarePairedDelimiter\abs{\lvert}{\rvert}
\DeclarePairedDelimiter\norm{\lVert}{\rVert}

\graphicspath{{figures/}}
\makeatletter
\def\input@path{{figures/}}
\makeatother

\begin{document}


\begin{abstract}
We introduce an algorithm that is simultaneously memory-efficient and low-scaling for applying ab initio molecular Hamiltonians to matrix-product states (MPS) via the tensor-hypercontraction (THC) format. These gains carry over to Krylov subspace methods, which can find low-lying eigenstates and simulate quantum time evolution while avoiding local minima and maintaining high accuracy. In our approach, the molecular Hamiltonian is represented as a sum of products of four MPOs, each with a bond dimension of only 2. Iteratively applying the MPOs to the current quantum state in MPS form, summing and re-compressing the MPS leads to a scheme with the same asymptotic memory cost as the bare MPS and reduces the computational cost scaling compared to the Krylov method using a conventional MPO construction. We provide a detailed theoretical derivation of these statements and conduct supporting numerical experiments to demonstrate the advantage. Our algorithm is highly parallelizable and thus lends itself to large-scale HPC simulations.
\end{abstract}

\maketitle

\section{Introduction}
\label{sec:intro}

We aim to simulate a molecular Hamiltonian, which is also known as electronic structure Hamiltonian, of the form (with $L$ the number of electronic spatial orbitals)
\begin{equation}
    H = T + V = \sum_{p,q=1}^L \sum_{\sigma \in \{\uparrow, \downarrow\}} t_{pq} \, a^{\dagger}_{p, \sigma} a_{q, \sigma} 
+ \frac{1}{2} \sum_{p,q,r,s=1}^L \sum_{\sigma, \sigma^{\prime} \in \{\uparrow, \downarrow\}} v_{pqrs} \, a^{\dagger}_{p, \sigma} a_{q, \sigma}a^{\dagger}_{r, \sigma^{\prime}} a_{s, \sigma^{\prime}}
\label{eq:ab_initio_H}
\end{equation}
using tensor network methods, specifically the matrix product state (MPS) formalism~\cite{PhysRevLett.75.3537, PhysRevB.55.2164, ma2022density}. $a^{\dagger}_{p, \sigma}$ and $a_{p, \sigma}$ are the fermionic creation and annihilation operators, respectively, and $t_{pq}$, $v_{pqrs}$ are coefficients resulting from single- and two-body orbital overlap integrals. To understand molecular properties such as the electronic structure~\cite{chan2011density, xu2022stochastic}, optoelectronic properties~\cite{ren2018time}, or molecular vibrations~\cite{baiardi2017vibrational}, the density matrix renormalization group (DMRG) method~\cite{PhysRevLett.69.2863, PhysRevB.48.10345} is widely applied to chemical systems with strong correlations, where traditional density functional theory and coupled cluster approaches face significant challenges~\cite{baiardi2017vibrational, xu2023new, chan2016matrix, keller2015efficient, SzalayOrsReview, Friesecke2022PredictingTF, ma2022density, cheng2024renormalized, xie2023kylin}. The development of attosecond-level experimental techniques~\cite{klunder2011probing, RevModPhys.81.163, lepine2014attosecond, goulielmakis2010real, nisoli2017attosecond, baltuvska2003attosecond, borrego2022attosecond, midorikawa2022progress} motivates the simulation of ultrafast electron dynamics since they determine the formation and breaking of chemical bonds~\cite{nisoli2017attosecond}. The time-dependent variational principle (TDVP) is a widely-used time evolution method to predict electron dynamics~\cite{PhysRevB.94.165116, Baiardi2019TheDM, baiardi2021electron, ren2022time, xu2022stochastic, xu2024kylin}. However, both of the above are variational methods in which the MPS evolves locally. It could result in the DMRG method getting trapped in local minima~\cite{Hu2015ExcitedStateGO, Baiardi2019TheDM} and might lead to an inaccurate time evolution simulation by TDVP~\cite{PhysRevB.97.024307, PhysRevB.99.054307} even for simple models~\cite{yang2020time}.     

In contrast, global Krylov subspace methods optimize all the sites globally and simultaneously~\cite{PAECKEL2019167998}, offering a reliable alternative method if DMRG or TDVP run into problems. Krylov subspace methods like the Lanczos algorithm~\cite{haydock1980recursive, lanczos1950iteration, dargel2012lanczos} can compute low-energy eigenstates reliably without local minima. The Lanczos algorithm also has the favorable capability of finding multiple excited states, being less sensitive to the results of lower eigenstates~\cite{wang2025improved, dektor2025inexact}. Conversely, using the DMRG algorithm, one has to explicitly project out the lower eigenstates, implying that inaccuracies propagate to the higher ones. To simulate time evolution, the global Krylov method~\footnote{There is also a ``local version'' for the Krylov method used in DMRG and time evolution simulation~\cite{PAECKEL2019167998}; in this paper, we solely focus on the global Krylov method introduced in Sec.~\ref{subsec:global_krylov}.} provides high-order error scaling~\cite{ronca2017time, frahm2019ultrafast, ren2022time} and works reliably. The value of the Krylov method lies in its ability to ensure accuracy while remaining robust across all models, rather than failing in a few cases like DMRG and TDVP methods.

However, the reachable system sizes and MPS bond dimensions in the Krylov methods are relatively small when using the molecular Hamiltonian in conventional matrix product operator (MPO) form. Especially when one chooses high-accuracy MPS compression methods such as singular value decomposition (SVD)~\cite{schollwock2011density, ren2022time}, the restriction results from the core step: applying the Hamiltonian to a quantum state, i.e., computing $H\ket{\psi}$ in the tensor network formalism and compressing it. Considering a molecular Hamiltonian of the form~\eqref{eq:ab_initio_H}, the maximum bond dimension $D$ scales as $\mathcal{O}(L^2)$ when using conventional MPO constructions~\cite{chan2016matrix, keller2015efficient, ren2020general}. One needs intensive memory to store $H\ket{\psi}$, whose bond dimension is the product of the MPS and MPO bond dimensions. Moreover, compressing $H\ket{\psi}$ to MPS form with smaller bond dimensions is essential for further calculations~\cite{frahm2019ultrafast, PAECKEL2019167998}; the computational cost is also high. The difficulty arises from the non-locality of the two-body integral tensor $v \in \mathbb{R}^{L \times L \times L \times L}$, which makes the molecular Hamiltonian more complicated than a Hamiltonian containing only local interactions.

\begin{figure}
\centering
\begin{tikzpicture}[scale = 0.6]
    
    \Vertex[x=1,label=\zeta,Math]{A}
    \Vertex[RGB,color={127,201,127}, shape = rectangle, x=3, y=1, label=\chi, Math]{B}
    \Vertex[RGB,color={127,201,127}, shape = rectangle, x=-1, y=1, label=\chi, Math]{C}
    \Vertex[RGB,color={127,201,127}, shape = rectangle, x=-1, y=-1, label=\chi, Math]{D}
    \Vertex[RGB,color={127,201,127}, shape = rectangle, x=3, y=-1, label=\chi, Math]{E}
    
    \Edge(A)(2.2, 0)
    \Edge(A)(-0.15, 0)
    \Edge(B)(2.2, 0)
    \Edge(B)(4, 2)
    \Edge(C)(-0.15, 0)
    \Edge(C)(-2.05, 2)
    \Edge(D)(-0.15, 0)
    \Edge(D)(-2.05, -2)
    \Edge(E)(2.2, 0)
    \Edge(E)(4, -2)

    \node at (1.8,0.24){$\nu$};
    \node at (0.2,0.24){$\mu$};

    \node at (-2.25, 2.1){$p$};
    \node at (-2.25, -2.1){$q$};
    \node at (4.25, 2.1){$r$};
    \node at (4.25, -2.1){$s$};

    \draw[dashed, blue] (-1.8,-1.8) rectangle (3.8,1.8);

    \Vertex[x=-7, y=0, shape = rectangle, ,size=0.7,color=green!50!blue]{F}
    \Edge(F)(-8.2, 1.2)
    \node at (-8.3, 1.3){$p$};
    \Edge(F)(-5.9, 1.2)
    \node at (-5.8, 1.4){$r$};
    \Edge(F)(-8.2, -1.2)
    \node at (-8.4, -1.4){$q$};
    \Edge(F)(-5.9, -1.2)
    \node at (-5.9, -1.4){$s$};
    
    \node at (-7, 0){$V$};

    \node at (-4, 0){$\approx$};
    
\end{tikzpicture}\
\caption{Graphical representation of the THC factorization to approximate the Coulomb (electron repulsion integral) tensor.}
\label{fig:THC}
\end{figure}

In this work, we propose and study an alternative Krylov method based on the tensor hypercontraction (THC) representation of $v$~\cite{hohenstein2012tensor, parrish2012tensor, hohenstein2012communication}:
\begin{equation}
v_{pqrs} \approx \sum ^ {N} _ {\mu,  \nu = 1} \chi^{\mu}_p  \chi^{\mu}_q \zeta^{\mu \nu}  \chi^{\nu}_r \chi^{\nu}_s,
\label{eq:thc0}
\end{equation}
where $N$ is the THC rank. This formulation involves only two distinct matrices $\chi$ and $\zeta$, as illustrated in Fig.~\ref{fig:THC}. We will show that the THC representation allows us to re-write the electronic Hamiltonian into a sum of sub-Hamiltonians, denoted THC-MPO, where each sub-Hamiltonian can be constructed as the product of four small MPOs with bond dimensions of only 2. Compared to calculations using a conventional MPO, such a small and constant bond dimension enables us to compute and compress $H\ket{\psi}$ with significantly reduced memory requirements and better complexity scaling; both are reduced by a factor of $\mathcal{O}(L^4)$ asymptotically. We demonstrate the advantages of our THC-MPO by utilizing it for low-lying eigenstates search and time evolution simulations based on Krylov subspace methods, exemplified by the water molecule \ce{H2O}, hydrogen chain with ten atoms \ce{H10}, and the Ammonia molecule \ce{NH3}. This allows us to track the accuracy and error sources by comparing them to results from the full configuration interaction (FCI) or exact diagonalization (ED) method. The numerical experiments show that our method enables us to calculate previously inaccessible system sizes when using Krylov methods with SVD compression; the memory advantages of our method become immediately apparent. Additionally, we will provide a general estimation of the computational complexity of larger molecules to highlight the potential. We will also illustrate why our method is well-suited for parallel computing.

\section{Theoretical background}
\label{sec:background}

\subsection{Matrix product states and operators}

In the tensor network framework, the wavefunction $\ket{\Psi}$ is typically represented as a matrix product state (MPS), also called tensor train~\cite{PhysRevLett.75.3537, PhysRevB.55.2164, schollwock2011density, Bridgeman_2017}:
\begin{equation}
    \ket{\Psi} =
    \sum_{n_1, \dots, n_L} A[1]^{n_1} A[2]^{n_2} \cdots A[L]^{n_L} \ket{n_1, \dots, n_L }.
\end{equation}
Each $A[i]$ is a tensor of order three, as shown in Fig.~\ref{fig:mps}. The superscript $n_i$ is a physical index enumerating the possible states at site $i$, and $A[i]^{n_i}$ is a $\chi_i \times \chi_{i+1}$ matrix for each $n_i$. The variable $\chi_i$ is the $i$-th bond dimension. We denote the maximum MPS bond dimension by $M$ in the following. 

\begin{figure}
\centering
\begin{subfigure}{0.45\textwidth}
    \centering
     \begin{tikzpicture}[scale=1, yshift=0.5]
    \Vertex[label=${A[1]}$, fontscale=0.9]{A}
    \Vertex[label=${A[2]}$, fontscale=0.9, x=1]{B}
    \Vertex[label=${A[3]}$, fontscale=0.9, x=2]{C}
    \Vertex[label=${A[L]}$, fontscale=0.9, x=4.8]{D}
    \Edge(A)(B)
    \Edge(B)(C)
    \Edge(A)(0,0.7)
    \Edge(B)(1,0.7)
    \Edge(C)(2,0.7)
    \Edge(C)(2.7,0)
    \Edge(D)(4.1,0)
    \Edge(D)(4.8,0.7)
    \draw[-][dotted](3, 0)--(3.8,0)[line width=1.5pt];
    \node at (-0.2,0.7)[scale=0.7]{$n_1$};
    \node at (0.8,0.7)[scale=0.7]{$n_2$};
    \node at (1.8,0.7)[scale=0.7]{$n_3$};
    \node at (4.6,0.7)[scale=0.7]{$n_{L}$};
\end{tikzpicture}
    \caption{MPS form of a quantum state.}
    \label{fig:mps}
\end{subfigure}
\hfill
\begin{subfigure}{0.45\textwidth}
    \centering
        \begin{tikzpicture}[scale=1] 
    \Vertex[label=${W[1]}$, fontscale=0.8, shape = rectangle,color=green!65!blue]{A}
    \Vertex[label=${W[2]}$, fontscale=0.8, x=1, shape = rectangle,color=green!65!blue]{B}
    \Vertex[label=${W[3]}$, fontscale=0.8, x=2, shape = rectangle,color=green!65!blue]{C}
    \Vertex[label=${W[L]}$, fontscale=0.8, x=4.8, shape = rectangle,color=green!65!blue]{D}

    \Edge(A)(0, -0.7)
    \Edge(B)(1, -0.7)
    \Edge(C)(2, -0.7)
    \Edge(D)(4.8, -0.7)

    \Edge(A)(B)
    \Edge(B)(C)

    \Edge(A)(0,0.7)
    \Edge(B)(1,0.7)
    \Edge(C)(2,0.7)
    \Edge(C)(2.7,0)
    \Edge(D)(4.1,0)
    \Edge(D)(4.8,0.7)

    \draw[-][dotted](3, 0)--(3.8,0)[line width=1.5pt];
    \node at (-0.25,0.7)[scale = 0.7]{$m_1$};
    \node at (0.75,0.7)[scale = 0.7]{$m_2$};
    \node at (1.75,0.7)[scale = 0.7]{$m_3$};
    \node at (4.5,0.7)[scale = 0.7]{$m_{L}$};
    \node at (-0.2, -0.7)[scale = 0.7]{$n_1$};
    \node at (0.8, -0.7)[scale = 0.7]{$n_2$};
    \node at (1.8, -0.7)[scale = 0.7]{$n_3$};
    \node at (4.6, -0.7)[scale = 0.7]{$n_{L}$};

\end{tikzpicture}
    \caption{MPO form of an operator.}
    \label{fig:mpo}
\end{subfigure}
\caption{Graphical tensor network representation of an MPS and MPO. The logical wavefunction and operator are obtained by contracting the matrices $A[i]^{n_i}$ and $W[i]^{m_i n_i}$, respectively.}
\end{figure}

Alongside MPS, there is a corresponding formalism for operators, known as matrix product operators (MPOs)~\cite{schollwock2011density, keller2015efficient, chan2016matrix}, given by:
\begin{equation}
\label{eq:MPO_def}
\hat{O} = \sum_{m,n} W[1]^{m_1 n_1} W[2]^{m_2 n_2} \ldots W[L]^{m_L n_L} \ket{m_1, \dots, m_L} \bra{n_1, \dots, n_L}.
\end{equation} 
Each element $W[i]^{m_i n_i}$ is a matrix of shape $\beta_i \times \beta_{i+1}$, where $\beta_i$ is the $i$-th bond dimension, as shown in Fig.~\ref{fig:mpo}. We denote the maximum MPO bond dimension by $D$ in the following. In Eq.~\eqref{eq:MPO_def}, an operator $\hat{O}$ is represented with respect to computational basis states $\ket{m_1, \dots, m_L} \bra{n_1, \dots, n_L} $, and the corresponding coefficients are obtained by contracting the $W$ matrices. 

Analogous to how applying a Hamiltonian matrix to a state vector yields a state vector, applying an MPO to an MPS will result in a new MPS by contracting the local physical tensor legs. Such an operation will increase the MPS bond dimension from $M$ to $D \cdot M$. Thus, it is significant to compress this resulting MPS for further calculations, especially for iterative applications appearing in Krylov methods. The combined application and compression procedure is illustrated in Fig.~\ref{fig:MPO_on_MPS}.

\begin{figure*}[t]
\centering
\begin{tikzpicture}[scale=0.8, >=stealth] 
\Vertex[label=${A[1]}$, fontscale=0.7,size=.5]{A}
\Vertex[label=${A[2]}$, fontscale=0.7, x=1,size=.5]{B}
\Vertex[label=${A[3]}$, fontscale=0.7,x=2,size=.5]{C}
\Vertex[label=${A[L]}$, fontscale=0.7,x=4.8,size=.5]{D}

\Edge(A)(B)
\Edge(B)(C)
\Edge(A)(0,0.7)
\Edge(B)(1,0.7)
\Edge(C)(2,0.7)
\Edge(C)(2.7,0)
\Edge(D)(4.1,0)
\Edge(D)(4.8,0.7)
\draw[-][dotted](3, 0)--(3.8,0)[line width=1.5pt];

\Vertex[label=${W[1]}$, fontscale=0.7, shape = rectangle, y = 1,color=green!65!blue,size=.5,size=.5]{E}
\Vertex[label=${W[2]}$, fontscale=0.7, x=1, y =1, shape = rectangle,color=green!65!blue,size=.5]{F}
\Vertex[label=${W[3]}$, fontscale=0.7, x=2, y =1,shape = rectangle,color=green!65!blue,size=.5]{G}
\Vertex[label=${W[L]}$, fontscale=0.7, x=4.8, y =1,shape = rectangle,color=green!65!blue,size=.5]{H}
\Edge(E)(0, 1.3)
\Edge(E)(0, 1.8)
\Edge(F)(1, 0)
\Edge(F)(1, 1.8)
\Edge(G)(2, 0)
\Edge(G)(2, 1.8)
\Edge(H)(4.8, 0)
\Edge(H)(4.8, 1.8)
\Edge(E)(F)
\Edge(F)(G)
\Edge(G)(2.7, 1)
\Edge(H)(4.1, 1)
\draw[-][dotted](3, 1)--(3.8, 1)[line width=1.5pt];

\node at (-0.3,1.7)[scale = 0.7]{$m_1$};
\node at (0.7,1.7)[scale = 0.7]{$m_2$};
\node at (1.7,1.7)[scale = 0.7]{$m_3$};
\node at (4.5,1.7)[scale = 0.7]{$m_L$};

\node at (2.7, 0.8)[scale = 0.7]{$\beta_4$};
\node at (2.7, 0.2)[scale = 0.7]{$\chi_4$};

\Edge[Direct](5.4,0.5)(6.7,0.5)

\node at (6, 0.8)[scale = 0.7]{Contract};

\Vertex[label=${A^\prime[1]}$, fontscale=0.7, x=7.4,size=.5]{I}
\Vertex[label=${A^\prime[2]}$, fontscale=0.7, x=8.4,size=.5]{J}
\Vertex[label=${A^\prime[3]}$, fontscale=0.7, x=9.4,size=.5]{K}
\Vertex[label=${A^\prime[L]}$, fontscale=0.65, x=12.2,size=.5]{L}
\Edge(I)(J)
\Edge(I)(7.4, 0.7)
\Edge(J)(K)
\Edge(J)(8.4, 0.7)
\Edge(K)(10.1, 0)
\Edge(K)(9.4, 0.7)
\draw[-][dotted](10.4, 0)--(11.2, 0)[line width=1.5pt];
\Edge(L)(11.4, 0)
\Edge(L)(12.2, 0.7)

\node at (10.1, 0.2)[scale = 0.7]{$\beta_4 \chi_4 $};

\node at (7.1,0.7)[scale = 0.7]{$m_1$};
\node at (8.1,0.7)[scale = 0.7]{$m_2$};
\node at (9.1,0.7)[scale = 0.7]{$m_3$};
\node at (11.9,0.7)[scale = 0.7]{$m_L$};

\Edge[Direct](12.7,0.5)(14,0.5)
\node at (13.3, 1.1)[scale = 0.7]{Canonical};
\node at (13.3, 0.8)[scale = 0.7]{-lize};
\node at (13.3, 0.2)[scale = 0.7]{Using QR};

\Vertex[style={shape=regular polygon, regular polygon sides=3, minimum size=1cm, rotate=90, , minimum size=0.7cm}, label=${B[1]}$, fontscale=0.6, x=14.7,size=.55]{M}
\Vertex[style={shape=regular polygon, regular polygon sides=3, minimum size=1cm, rotate=90, , minimum size=0.7cm}, label=${B[2]}$, fontscale=0.6, x=15.7,size=.55]{N}
\Vertex[style={shape=regular polygon, regular polygon sides=3, minimum size=1cm, rotate=90, , minimum size=0.7cm}, label=${B[3]}$, fontscale=0.6, x=16.7,size=.55]{O}
\Vertex[style={shape=regular polygon, regular polygon sides=3, minimum size=1cm, rotate=90, , minimum size=0.7cm}, label=${B[L]}$, fontscale=0.55, x=19.5,size=.57]{P}

\Edge(M)(N)
\Edge(M)(14.7, 0.7)
\Edge(N)(O)
\Edge(N)(15.7, 0.7)
\Edge(O)(17.4, 0)
\Edge(O)(16.7, 0.7)
\draw[-][dotted](17.7, 0)--(18.5, 0)[line width=1.5pt];
\Edge(P)(18.7, 0)
\Edge(P)(19.5, 0.7)

\node at (14.4,0.7)[scale = 0.7]{$m_1$};
\node at (15.4,0.7)[scale = 0.7]{$m_2$};
\node at (16.4,0.7)[scale = 0.7]{$m_3$};
\node at (19.2,0.7)[scale = 0.7]{$m_L$};

\Edge[Direct](-0.2, -2)(1,-2)
\node at (0.3, -1.7)[scale = 0.7]{SVD};

\Vertex[style={shape=regular polygon, regular polygon sides=3, minimum size=1cm, rotate=-90, minimum size=0.7cm}, label=${U[1]}$, fontscale=0.6, x=1.4, y=-2, size=.5]{Q}
\Vertex[label=${S}$, fontscale=0.7, x=2.4, y=-2, size=.5]{R}
\Vertex[style={shape=regular polygon, regular polygon sides=3, minimum size=1cm, rotate=90, minimum size=0.7cm}, label=${V^{\dagger}}$, fontscale=0.6, x=3.4, y=-2, size=.5]{S}
\Vertex[style={shape=regular polygon, regular polygon sides=3, minimum size=1cm, rotate=90, minimum size=0.7cm}, label=${B[2]}$, fontscale=0.6, x=4.4, y=-2, size=.5]{T}

\Vertex[style={shape=regular polygon, regular polygon sides=3, minimum size=1cm, rotate=90, minimum size=0.7cm}, label=${B[3]}$, fontscale=0.6, x=5.4, y=-2, size=.5]{V}

\Vertex[style={shape=regular polygon, regular polygon sides=3, minimum size=1cm, rotate=90, minimum size=0.7cm}, label=${B[L]}$, fontscale=0.6, x=8.6, y=-2, size=.5]{U}

\draw[-][dotted](1.1, -2.5)--(3.75, -2.5)[line width=1pt];
\draw[-][dotted](1, -2.5)--(1, -1.55)[line width=1pt];
\draw[-][dotted](3.75, -1.55)--(1.05, -1.55)[line width=1pt];
\draw[-][dotted](3.75, -1.65)--(3.75, -2.5)[line width=1pt];

\Edge[color=red](Q)(R)
\Edge(Q)(1.4, -1.3)
\Edge[color=red](R)(S)

\Edge(S)(T)
\Edge(V)(T)

\Edge(T)(5.1, -2) 
\Edge(T)(4.4, -1.3)
\draw[-][dotted](6.4, -2)--(7.2, -2)[line width=1.5pt];
\Edge(U)(8.4, -2)
\Edge(U)(8.6, -1.3)

\Edge(V)(5.4, -1.3) 
\Edge(V)(6.1, -2) 
\Edge(U)(7.7, -2)

\node at (1.1,-1.3)[scale = 0.7]{$m_1$};
\node at (4.1,-1.3)[scale = 0.7]{$m_2$}; 
\node at (5.1,-1.3)[scale = 0.7]{$m_3$}; 
\node at (8.3,-1.3)[scale = 0.7]{$m_L$};

\draw[decorate, decoration={brace, amplitude=3pt, mirror, raise=1pt}, blue]
    (2.4, -2.5) -- (3.4, -2.5) node [midway, yshift=-10pt, text=green!30!blue] {};

\draw[->, blue] (2.9, -2.75) .. controls (3.4, -3) and (3.7, -2.95) .. (4.2, -2.5);

\node at (3.6, -3.1)[scale = 0.7]{\textbf{\texttt{Absorb}}};
\draw[->] (2.4,-1.65)-- (2.6,-1.2);
\node at (2.6,-1)[scale = 0.7]{\textbf{\texttt{Truncate}}};

\Edge[Direct](9.4, -2)(10.6,-2)
\node at (9.9, -1.7)[scale = 0.7]{Truncate};
\node at (9.9, -2.3)[scale = 0.7]{Absorb};

\Vertex[style={shape=regular polygon, regular polygon sides=3, minimum size=1cm, rotate=30, minimum size=0.7cm}, label=${U[1]}$, fontscale=0.6, x=12.1, y=-2, size=.6]{M}
\Vertex[label=${B^{\prime}[2]}$, fontscale=0.55, x=13.25, y=-2, size=.5]{N}

\Vertex[style={shape=regular polygon, regular polygon sides=3, minimum size=1cm, rotate=90, minimum size=0.7cm}, label=${B[3]}$, fontscale=0.6, x=14.25, y=-2, size=.6]{O}
\Vertex[style={shape=regular polygon, regular polygon sides=3, minimum size=1cm, rotate=90, minimum size=0.7cm}, label=${B[L]}$, fontscale=0.55, x=17.05, y=-2, size=.6]{P}

\Edge(M)(N)
\Edge(M)(12.1, -1.3)
\Edge(N)(O)
\Edge(N)(13.25, -1.3)
\Edge(O)(14.95, -2)
\Edge(O)(14.25, -1.3)
\draw[-][dotted](15.25, -2)--(16.05, -2)[line width=1.5pt];
\Edge(P)(16.25, -2)
\Edge(P)(17.05, -1.3)

\node at (11.8,-1.3)[scale = 0.7]{$m_1$};
\node at (12.9,-1.3)[scale = 0.7]{$m_2$};
\node at (13.9,-1.3)[scale = 0.7]{$m_3$};
\node at (16.7,-1.3)[scale = 0.7]{$m_L$};

\Edge[Direct](17.7, -2)(18.8,-2)
\node at (18.2, -1.7)[scale = 0.7]{2nd site};
\node at (18.2, -2.25)[scale = 0.7]{SVD};
\draw[-][dotted](19,-2)--(19.4,-2)[line width=1.5pt];
\end{tikzpicture}
\caption{Multiplying an MPO with an MPS and subsequent compression. We first contract the tensors along the physical axis. Then, the MPS is transformed into right-canonical form by QR decompositions. Next, we employ SVDs from left to right to reduce the bond dimension by discarding the smallest singular values and merging $S$ and $V^{\dagger}$ matrices into the next site.}
\label{fig:MPO_on_MPS}
\end{figure*}

\subsection{The Krylov methods based on matrix product states}
\label{subsec:global_krylov}

The complexity of the Hamiltonian in full matrix form scales exponentially with system size; thus, exact diagonalization approaches are restricted to relatively small systems. A possibility to overcome this restriction consists in combining the Lanczos algorithm~\cite{haydock1980recursive, lanczos1950iteration} with the MPS representation~\cite{dargel2011adaptive, dargel2012lanczos}. Namely, the MPS ansatz offers an economical representation of quantum states, and the Lanczos algorithm confines the evolving space to the Krylov subspace spanned by $\{ \ket{\psi}, H\ket{\psi}, H^2 \ket{\psi},\dots, H^{K-1}\ket{\psi} \}$. In this work, we construct an orthogonal basis of the Krylov subspace. Starting from some initial state $\ket{v_0} \equiv \ket{\psi}$, we compute the next Krylov vector $\ket{v_{i+1}}$ by applying the Hamiltonian to $\ket{v_{i}}$ and orthogonalizing it w.r.t.\ the previous ones using the Gram-Schmidt algorithm~\cite{frahm2019ultrafast, PAECKEL2019167998, dargel2012lanczos}. It is also possible to use these Krylov vectors without orthogonalization; see~\cite{frahm2019ultrafast, PAECKEL2019167998} for further details.

The Hamiltonian is projected onto the Krylov subspace, and the elements of the resulting effective Hamiltonian are given by
\begin{equation}
\widetilde{H}_{ij} = \bra{v_i}H \ket{v_j},
\label{eq:Krylov_H}
\end{equation}
where $\{\ket{v_0}, \ket{v_1}, \dots,  \ket{v_{K-1}} \}$ are orthonormalized Krylov vectors forming a basis of the subspace. Typically, we assume this effective Hamiltonian to be tridiagonal so that we only retain the entries with $\abs{i - j} \le 1$~\cite{dargel2012lanczos, PAECKEL2019167998}. Assuming that the subspace dimension $K$ is sufficiently large, the diagonalization of this effective Hamiltonian $\widetilde{H}$ provides a reliable estimate of the low-lying eigenstates. Such a procedure is the well-known Lanczos algorithm~\cite{haydock1980recursive, lanczos1950iteration, dargel2012lanczos}. Regarding the Hamiltonian as a linear combination of eigenspace projectors, the Krylov space can only contain components already present in the initial state. Therefore, one can also obtain excited states starting from an initial state orthogonal to the lower eigenstates. 

Time evolution can also be simulated based on the Krylov subspace~\cite{PAECKEL2019167998,park1986unitary, kosloff1988time, frahm2019ultrafast, hochbruck1997krylov}, where the Krylov vectors $\{\ket{v_0}, \ket{v_1}, \dots, \ket{v_{K-1}} \}$ are built based on the initial state at $t = 0$. A general quantum state in such a subspace can be written as:
\begin{equation}
    \ket{\Psi} = \sum_{i=0}^{K-1} a_i \ket{v_i}.
\label{eq:Krylov_combination}
\end{equation}
We represent the state as $\Vec{a} = (a_0, a_1, \dots, a_{K-1})^T$, where $a_i$ refers to the amplitude for each basis vector. With the help of such a formalism, the time-evolved state in the Krylov subspace is formulated as:
\begin{equation}
    \Vec{a}(t=\delta) = e^{-i\delta \widetilde{H}}\Vec{a}(t=0),
\end{equation}
where $\Vec{a}(t=0)$ is $(1, 0, 0, \dots, 0)^T$ since the initial state is just $\ket{v_0}$. One can explicitly reconstruct the time-evolved quantum state by Eq.~\eqref{eq:Krylov_combination}. The accuracy of this time evolution method depends on the size of the Krylov subspace, and the error is of order $\mathcal{O}(\delta^K)$ for a single time step $\delta$ and thus $\mathcal{O}(\delta^{K-1})$ for a fixed duration~\cite{PAECKEL2019167998}.  The accuracy varies for different models: the upper error bounds are determined by the spectral width $W$, the step size $\delta$, and the subspace size $N$;
see Appendix~\ref {app:time_evo_error} for quantitative discussion. As a practical guide, a subspace dimension of 3 $\sim$ 10 is typically sufficient to achieve satisfactory accuracy when selecting small time step sizes, as suggested in~\cite{PAECKEL2019167998}.

When using the MPS formalism to implement these algorithms, extra errors are introduced due to the MPS truncation, particularly the loss of orthogonality of the Krylov basis. We employ the strategy proposed in~\cite{dargel2012lanczos} to address this issue, and we noted that the canonical orthogonalization method might also be useful~\cite{jiang2021chebyshev}. Both these techniques aim at finding a linear combination:
\begin{equation}
    \ket{\psi_a} = \sum_{i = 0}^{a} C_{ai} \ket{v_i},
\label{eq:reortho}
\end{equation} of current Krylov vectors $\ket{v_i}$, so that the resulting vectors $\ket{\psi_a}$ are well-orthogonalized. Note that we do not change the $\ket{v_i}$ vectors; instead, we only focus on solving the matrix $C$. Consequently, the elements of effective Hamiltonian in the $\ket{\psi_a}$ basis are given by:
\begin{equation}
    \Tilde{H}_{ab} = \bra{\psi_a} H \ket{\psi_b} = \sum_{i = 0}^{a} \sum_{j = 0}^{b} C^{*}_{ai} C_{bj} \bra{v_i} H \ket{v_j}.
\label{eq:eff_H}
\end{equation}
The MPS truncation will still reduce accuracy even though the Krylov vectors are well orthogonalized; we find that restarting the Lanczos algorithm is helpful in improving the convergence. For simulating time evolution, truncation errors become significant only at very small time steps.

The Krylov method's most expensive and memory-intensive part is obtaining the Krylov vectors whose core step is to compute $H \ket{v_i}$. Conventionally, one multiplies the Hamiltonian's MPO with an MPS, resulting in an intermediate MPS with a large maximum bond dimension $\mathcal{O}(L^2 M)$ for ab initio molecular Hamiltonians. The memory cost to store such an intermediate MPS scales as $\mathcal{O}(L^5 M^2)$ for all sites, which could exhaust the available memory for even small-sized systems. Second, the high computational cost of compressing the intermediate MPS of $H\ket{v_i}$ is another bottleneck of constructing the Krylov subspace. One must compress the bond dimensions of $H\ket{v_i}$ back to smaller target bond dimensions to avoid the exponential increase in the next multiplications. To achieve fully controllable and highly accurate truncation, one typically employs the SVD method. In this approach, one first brings the intermediate MPS into canonical form using QR decomposition and then truncates the bonds by SVD, as depicted in Fig.~\ref{fig:MPO_on_MPS}. Given that one can easily read off the Schmidt values from the mixed-canonical form, the truncation can be performed with a desired accuracy~\cite{hauschild2018efficient, schollwock2011density}. The QR decompositions are the main contributors to the cost of compression. The computational complexity of QR decomposition for an MPS tensor with bond dimension $\mathcal{O}(L^2M)$ is $\mathcal{O}(L^6 M^3)$, leading to a total complexity across all sites as high as $\mathcal{O}(L^7 M^3)$ which makes it challenging to apply the Krylov method on large molecules. 

There are also alternative MPS compression methods. The zip-up method~\cite{Stoudenmire_2010} is more efficient, but since the algorithm works on a non-orthogonalized basis, the error is not fully controlled. The variational method requires a proper initial guess. Otherwise, one needs a large number of iterations and sweeps~\cite{schollwock2011density}. The recently proposed density matrix method~\cite{ma2024approximate} provides another fully controllable compression scheme that merits further study in future research. Our THC-MPO discussed in this paper can improve most of these MPS compression schemes when simulating molecular Hamiltonians, see Appendix~\ref{app:alternative} for more details; we focus on the traditional SVD method in this paper. We would like to emphasize that our method is compatible with all the MPS compression methods, and we can always check whether our method can be integrated into them when a newer compression scheme is proposed.

\subsection{The THC factorization}
\label{sec:THC}

Employed widely in the simulation of molecular systems already~\cite{lee2019systematically, hohenstein2012communication, parrish2014communication, schutski2017tensor, lee2019systematically}, the tensor hypercontraction (THC) proposed by E.~Hohenstein et al.~\cite{lee2019systematically, hohenstein2012communication, parrish2014communication} approximates the two-electron integrals $v_{pqrs}$ as:
\begin{equation}
    v_{pqrs} \approx \sum ^ {N} _ {\mu,  \nu = 1} \chi^{\mu}_p  \chi^{\mu}_q \zeta^{\mu \nu}  \chi^{\nu}_r \chi^{\nu}_s
\label{eq:thc}
\end{equation}
for all $p, q, r, s \in \{ 1, \dots, L \}$, as illustrated in Fig.~\ref{fig:THC}.

Currently, several relatively mature algorithms exist to obtain these tensors. The original papers proposed the PF-THC~\cite{hohenstein2012tensor} and LS-THC~\cite{parrish2012tensor} methods as algorithms. Subsequently, the interpolative separable density fitting (ISDF) method~\cite{LU2015329} enhanced the computational efficiency and improved the approximation accuracy. In density-fitting (DF)~\cite{whitten1973coulombic, dunlap1979some, werner2003fast}, one approximates the product of two orbitals as:
\begin{equation}
\rho_{pq}(r) \coloneqq \phi_p (r) \phi_{q}(r) \approx \sum_{\mu=1}^{N_{a}} C^{\mu}_{pq} P_{\mu}(r),
\end{equation}
where $P_{\mu}$ for $\mu = 1, 2, \dots, N_{a}$ are auxiliary basis functions. The idea of ISDF is that if we approximate $\rho_{pq}$ by interpolation, the THC factorization can directly be obtained~\cite{LU2015329}:
\begin{equation}
\rho_{pq}(r) \approx \sum_k \rho_{pq}(r_k) F_{k}(r) = \sum_k \phi_p (r_k) \phi_{q}(r_k) F_{k}(r),
\end{equation}
where $r_k$ are selected grid points in the Becke scheme. The selection is implemented by interpolative decomposition, aimed at choosing a limited number of rows to approximate $\rho_{pq}(r_k)$ interpreted as a $N_g \times L^2$ matrix, where $N_g$ is the total number of Becke grid points. Since the row indices represent individual grid points, the procedure can also be interpreted as discarding less important grid points. Their importance is revealed by randomized QR decomposition with column-pivoting~\cite{LU2015329, hu2017interpolative}. We then determine fit functions $F_{k}$ after we obtain selected grid points. The fit functions are chosen as auxiliary basis functions $P_{\mu}$ in~\cite{LU2015329}, but in this work, we obtained them following the strategy introduced in LS-THC~\cite{parrish2012tensor}, as suggested in~\cite{lee2019systematically}. 


To improve the accuracy of the THC decomposition, we minimize the relative error:
\begin{equation}
    \epsilon _V = \frac{\| v_{pqrs} - \sum ^ {N} _ {\mu,  \nu = 1} \chi^{\mu}_p  \chi^{\mu}_q \zeta^{\mu \nu}  \chi^{\nu}_r \chi^{\nu}_s  \|
}{\|v_{pqrs}\|}.
\end{equation}
where $\|v_{pqrs}\| = \sqrt{\sum_{pqrs} \left| v_{pqrs} \right|^2}$ denotes the Frobenius norm. It is essentially an optimization problem, and we carried it out using the Adam optimizer~\cite{kingma2015adam}, implemented in Optax~\cite{optax}. Although Adam was introduced for stochastic optimization problems, its adaptive moment mechanism also converges reliably in deterministic settings. In our tests, Adam converged faster than classical optimizers such as the Broyden-Fletcher-Goldfarb-Shanno (BFGS) algorithm~\cite{nocedal2006numerical}, so we adopted it for the optimization step. It appears that there are two matrices, namely $\chi$ and $\zeta$, to be optimized. However, since $\zeta$ can be obtained from $\chi$ by LS-THC, the number of free parameters is reduced to the entries of $\chi$. Exemplified by the hydrogen chain, we first carry out the optimization by 1000 rounds with a learning rate of 0.001, followed by another 1000 rounds with a learning rate of 0.0005. In the numerical experiment, we reach the acceptable chemical accuracy (1.6 mHatree) with $N = 4L$ for the water molecule \ce{H2O}, $N = 3L -3$ for the hydrogen chain \ce{H10}, and $N = 4.5L$ for the Ammonia molecule \ce{NH3}, all in the STO-6G basis set. We ensure the accuracy reaches chemical accuracy by comparing it with the energy obtained via FCI results from PySCF. Regarding our future studies in larger systems, we noticed that previous studies have demonstrated that the THC rank $N$ typically exhibits near-linear scaling with respect to the system size $L$~\cite{hohenstein2012tensor, parrish2012tensor, hohenstein2012communication, PRXQuantum.2.030305, matthews2020improved, lee2019systematically, LU2015329, hu2017interpolative, dong2018interpolative}. For instance, there are 76 spatial orbitals needed for the active-space model of the FeMoco system proposed by Li et al.~\cite{li2019electronic}, and the THC rank of 450 is enough to reach the chemical accuracy~\cite{PRXQuantum.2.030305}; for hydrogen chain of $L$ atoms ($L$ spatial orbitals in STO-6G basis) with distances of $1.4$ Angstrom, a THC rank of $3L-3$ is sufficient to achieve the accuracy of $5 \times 10^{-5}$ Hartree per atom~\cite{luo2025efficient}. We remark that this part is a pre-processing step. If a prepared THC decomposition is available (e.g., from other work or a database), we can apply our THC-MPO method directly.

\section{The Krylov method based on THC}

\subsection{Constructing MPOs using THC}

In this section, we first show how to use the THC factorization to construct a special representation of the molecular Hamiltonian (THC-MPO). Then, we will utilize the THC-MPO in Krylov methods and discuss its advantages. We focus on the challenging Coulomb term here. An MPO of the kinetic term $T$ can be easily constructed following the strategy introduced in~\cite{chan2016matrix}, and we will also discuss the kinetic term in Sec.~\ref{subsec:thc-krylov}.

Inserting the THC factorization Eq.~\eqref{eq:thc} into the Coulomb term $V$, one immediately arrives at:
\begin{equation}
V \approx \frac{1}{2} \sum_{\mu, \nu = 1}^N \sum _{\sigma, \sigma^{\prime}\in \{\uparrow, \downarrow\}} G_{\mu \sigma, \nu \sigma^{\prime}},   
\label{eq:sum_of_H}
\end{equation}
where $G_{\mu \sigma, \nu \sigma^{\prime}}$ is defined as:
\begin{equation}
G_{\mu \sigma, \nu \sigma^{\prime}} =  
\zeta^{\mu \nu}
\left( \sum_{p = 1}^{L} \chi_p^{\mu} a^{\dagger}_{p,\sigma} \right)
\left( \sum_{q = 1}^{L} \chi_q^{\mu} a_{q,\sigma} \right)
\left( \sum_{r = 1}^{L} \chi_r^{\nu} a^{\dagger}_{r,\sigma^{\prime}} \right)
\left( \sum_{s = 1}^{L} \chi_s^{\nu} a_{s,\sigma^{\prime}} \right)
\label{eq:G_mu_nu_def}
\end{equation}

Each sub-term (exemplified by $\sum^{L}_s \chi^{\nu}_s a_{s,\sigma^{\prime}}$) in $G_{\mu \sigma, \nu \sigma^{\prime}}$ can explicitly be converted to an MPO as follows:
\begin{subequations}
\label{eq:elementary_MPO}
\begin{equation}
W[s] = 
\begin{pmatrix}
I & \chi^{\nu}_s a_{s,\sigma^{\prime}}\\
0 & I \\
\end{pmatrix}, \quad s = 2, \dots, L-1
\label{eq:singleW}
\end{equation}
and the first and last tensors:
\begin{equation}
W[1] = \begin{pmatrix}
    I & \chi^{\nu}_1 a_{1,\sigma^{\prime}} 
\end{pmatrix}, \quad
W[L] = \begin{pmatrix}
    \chi^{\nu}_L a_{L,\sigma^{\prime}}\\
    I
\end{pmatrix}.
\end{equation}
\end{subequations}
One can contract the $W$ matrices sequentially to verify the correctness of the construction. It is worth noting that the bond dimension of $W[s]$ is always only $2$, independent of the system size $L$. 

In this work, we follow the convention of treating each spatial orbital as a single site in the MPS. When implementing a corresponding MPO numerically using the Jordan-Wigner transformation~\cite{jordan1928equivalenzverbot}, we replace the fermionic operators with their bosonic counterparts and substitute the identities in each $W$ at position $(1,1)$ by Pauli-$Z$ operators (to account for fermionic sign factors):
\begin{equation}
W[s] = \begin{pmatrix}
Z\otimes Z & \chi^{\nu}_s b_{s,\sigma^{\prime}}\\
0 & I_4 \\
\end{pmatrix},
\end{equation}
where $I_4$ denotes the identity matrix of size $4 \times 4$, and $b_{s,\sigma^{\prime}}$ is defined as the local annihilation operator for spin $\sigma^{\prime}$ of size $4 \times 4$, for which we detail these in Appendix~\ref{app:local_op}.

Following the strategy above, one can analogously construct MPOs for the other three sub-terms: $\sum^{L}_{p} \chi^{\mu}_p a^{\dagger}_{p,\sigma}$, $\sum^{L}_{q} \chi^{\mu}_q a_{q,\sigma}$ and $\sum^{L}_{r} \chi^{\nu}_r a^{\dagger}_{r,\sigma^{\prime}}$. The entire MPO of $G_{\mu \sigma, \nu \sigma^{\prime}}$ is thus the product of the MPOs of these four sub-terms as shown in Fig.~\ref{fig:thc_mpo}, and the scalar $\zeta^{\mu \nu}$ can be absorbed into $W[1]$ at the first site. Therefore, the MPO of $G_{\mu \sigma, \nu \sigma^{\prime}}$ likewise has a constant bond dimension. The whole MPO of the Coulomb term is thus the summation of MPOs of sub-Hamiltonians $G_{\mu \sigma, \nu \sigma^{\prime}}$, but to calculate $V \ket{\psi}$ in compressed MPS form, we will refrain from merging them into a large MPO, see below.

\begin{figure}
\centering
\begin{tikzpicture}[scale=0.75] 
    \Vertex[shape = rectangle, color=green!80!blue]{A}
    \Vertex[x=1, shape = rectangle, color=green!80!blue]{B}
    \Vertex[x=2, shape = rectangle, color=green!80!blue]{C}
    \Vertex[x=4.8, shape = rectangle, color=green!80!blue]{D}
    \Vertex[x=5.8, shape = rectangle, color=green!80!blue]{E}
    
    \Edge(A)(B) \Edge(B)(C) \Edge(D)(E)
    \Edge(A)(0, -0.6) \Edge(B)(1, -0.6)
    \Edge(C)(2, -0.6) \Edge(D)(4.8, -0.6)
    \Edge(E)(5.8, -0.6)
    
    \Edge(A)(0,0.6) \Edge(B)(1,0.6)
    \Edge(C)(2,0.6) \Edge(C)(2.7,0)
    \Edge(D)(4.1,0) \Edge(D)(4.8,0.6)
    \Edge(E)(5.8,0.6)

    \draw[-][dotted](3, 0)--(3.8,0)[line width=1.5pt];

    \Vertex[x=0, y=1.4, shape = rectangle, color=green!65!blue]{F}
    \Vertex[x=1, y=1.4, shape = rectangle, color=green!65!blue]{G}
    \Vertex[x=2, y=1.4, shape = rectangle, color=green!65!blue]{H}
    \Vertex[x=4.8,y=1.4, shape = rectangle, color=green!65!blue]{I}
    \Vertex[x=5.8,y=1.4, shape = rectangle, color=green!65!blue]{J}

    \Edge(F)(G) \Edge(G)(H) \Edge(H)(2.7,1.4)
     \Edge(I)(4.1,1.4) \Edge(I)(J)
    
    \Edge(F)(0, 0.8) \Edge(G)(1, 0.8) 
    \Edge(H)(2, 0.8) \Edge(I)(4.8, 0.8)
    \Edge(J)(5.8, 0.8)

    \Edge(F)(0,2) \Edge(G)(1,2) \Edge(H)(2,2)
    \Edge(I)(4.8,2) \Edge(J)(5.8,2)
    
    \draw[-][dotted](3, 1.4)--(3.8,1.4)[line width=1.5pt];

    \Vertex[x=0, y=2.8, shape = rectangle, color=green!50!blue]{K}
    \Vertex[x=1, y=2.8, shape = rectangle, color=green!50!blue]{L}
    \Vertex[x=2, y=2.8, shape = rectangle, color=green!50!blue]{M}
    \Vertex[x=4.8,y=2.8, shape = rectangle, color=green!50!blue]{N}
    \Vertex[x=5.8,y=2.8, shape = rectangle, color=green!50!blue]{O}

    \Edge(K)(L) \Edge(L)(M) \Edge(M)(2.7,2.8)
    \Edge(N)(4.1,2.8) \Edge(N)(O)

    \Edge(K)(0, 2.2) \Edge(L)(1, 2.2)
    \Edge(M)(2, 2.2) \Edge(N)(4.8, 2.2)
    \Edge(O)(5.8, 2.2)

    \Edge(K)(0,3.4) \Edge(L)(1,3.4) \Edge(M)(2,3.4)
    \Edge(N)(4.8,3.4) \Edge(O)(5.8,3.4)

    \draw[-][dotted](3, 2.8)--(3.8,2.8)[line width=1.5pt];

    \Vertex[x=0, y=4.2, shape = rectangle, color=green!35!blue]{P}
    \Vertex[x=1, y=4.2, shape = rectangle, color=green!35!blue]{Q}
    \Vertex[x=2, y=4.2, shape = rectangle, color=green!35!blue]{R}
    \Vertex[x=4.8,y=4.2, shape = rectangle, color=green!35!blue]{S}
    \Vertex[x=5.8,y=4.2, shape = rectangle, color=green!35!blue]{T}

    \Edge(P)(Q) \Edge(Q)(R) \Edge(R)(2.7,4.2)
    \Edge(S)(4.1,4.2) \Edge(S)(T)

    \Edge(P)(0, 3.6) \Edge(Q)(1, 3.6)
    \Edge(R)(2, 3.6) \Edge(S)(4.8, 3.6)
    \Edge(T)(5.8, 3.6)

    \Edge(P)(0,4.8) \Edge(Q)(1,4.8) \Edge(R)(2,4.8)
    \Edge(S)(4.8,4.8) \Edge(T)(5.8,4.8)

    \draw[-][dotted](3, 4.2)--(3.8,4.2)[line width=1.5pt];

    \node at (-2.8,0)[scale=1]{$\sum^{L}_{s=1} \chi^{\nu}_s a_{s,\sigma^{\prime}}$};
    \node at (-2.8,1.4)[scale=1]{$\sum^{L}_{r=1} \chi^{\nu}_r a^{\dagger}_{r,\sigma^{\prime}}$};
    \node at (-2.8,2.8)[scale=1]{$\sum^{L}_{q=1} \chi^{\mu}_q a_{q, \sigma}$};
    \node at (-2.8,4.2)[scale=1]{$\sum^{L}_{p=1} \chi^{\mu}_p a^{\dagger}_{p,\sigma}$};
    \Edge[Direct](-1.5,0)(-0.7, 0)
    \Edge[Direct](-1.5,1.4)(-0.7, 1.4)
    \Edge[Direct](-1.5,2.8)(-0.7, 2.8)
    \Edge[Direct](-1.5,4.2)(-0.7, 4.2)

    \Vertex[fontscale=0.9, y = -2, color= orange,size=.57]{U}
    \Vertex[fontscale=0.9, x=1, y = -2, color= orange,size=.57]{V}
    \Vertex[fontscale=0.9,x=2, y = -2, color= orange,size=.57]{W}
    \Vertex[fontscale=0.9,x=4.8, y = -2, color= orange,size=.57]{X}
    \Vertex[fontscale=0.9,x=5.8, y = -2, color= orange,size=.57]{Y}
    
    \Edge(U)(V) \Edge(V)(W) \Edge(X)(Y)
    \Edge(U)(0, -1.4) \Edge(V)(1, -1.4) \Edge(W)(2, -1.4)
    \Edge(X)(4.8, -1.4) \Edge(Y)(5.8, -1.4)
    \Edge(W)(2.7, -2) \Edge(X)(4.1, -2)
    \draw[-][dotted](3,-2)--(3.8,-2)[line width=1.5pt];

    \draw[-][dotted](-0.6,-2.5)--(6.4,-2.5)[line width=0.8pt];
    \draw[-][dotted](-0.6,-2.5)--(-0.6,0.47)[line width=0.8pt];
    \draw[-][dotted](6.4,0.47)--(-0.6,0.47)[line width=0.8pt];
    \draw[-][dotted](6.4,0.47)--(6.4,-2.5)[line width=0.8pt];

    \Edge[Direct](3.2, -0.8)(3.2, -1.3)
    \Edge[Direct](3.2, 1.05)(3.2, 0.55)
    \Edge[Direct](3.2, 2.3)(3.2, 1.8)
    \Edge[Direct](3.2, 3.7)(3.2, 3.2)
\end{tikzpicture}
\caption{$G_{\mu \sigma, \nu \sigma^{\prime}}$ in Eq.~\eqref{eq:G_mu_nu_def} is composed of four layers of MPOs (square tensors) with bond dimension $2$. As specified by the arrows, we contract and compress the layers one at a time with the MPS (orange tensors).}
\label{fig:thc_mpo}
\end{figure}

\subsection{Krylov method using THC-MPO}
\label{subsec:thc-krylov}

As discussed in Sec.~\ref{subsec:global_krylov}, the essential step in Krylov methods is multiplying $H$ with $\ket{\psi}$. Here, we focus on the Coulomb term $V$ in $H = T + V$. We present how to take advantage of our THC-MPO to execute the multiplication and subsequent compression. With the help of Eq.~\eqref{eq:sum_of_H}, we can write $V\ket{\psi}$ as:
\begin{equation}
V \ket{\psi} \approx \frac{1}{2} \sum_{\mu, \nu = 1}^N \sum _{\sigma, \sigma^{\prime}\in \{\uparrow, \downarrow\}} G_{\mu \sigma, \nu \sigma^{\prime}} \ket{\psi},
\label{eq:H_on_mps_subterms}
\end{equation}
where we apply each sub-Hamiltonian to $\ket{\psi}$ and sum the resulting states up. Instead of manipulating large matrices, we compute $V \ket{\psi}$ via the small MPOs of $G_{\mu \sigma, \nu \sigma^{\prime}}$. 

For each term $G_{\mu \sigma, \nu \sigma^{\prime}} \ket{\psi}$, we execute multiplication and compression for each elementary MPO (layers shown in Fig.~\ref{fig:thc_mpo}) sequentially instead of treating $G_{\mu \sigma, \nu \sigma^{\prime}}$ as a whole. Each compression returns the bond dimension to $M$ so that the maximum bond dimension is only $2M$ during the calculation (since the MPO bond dimension for each layer is $2$). Beginning with $G_{1\uparrow, 1\uparrow} \ket{\psi}$, we add each subsequent $G_{\mu \sigma, \nu \sigma^{\prime}} \ket{\psi}$. Such an MPS addition likewise leads to an intermediate MPS of bond dimension $2M$, which is still cheap to store and compress. In summary, $\mathcal{O}(L M^2)$ memory is required to store the largest intermediate MPS, which is less by a factor $\mathcal{O}(L^4)$ compared to a conventional MPO algorithm. In addition, the memory for storing $G_{\mu \sigma, \nu \sigma^{\prime}} \ket{\psi}$ and $G_{\mu \sigma, \nu \sigma^{\prime}}$ is immediately released after adding $G_{\mu \sigma, \nu \sigma^{\prime}} \ket{\psi}$ to others. Implementing Eq.~\eqref{eq:H_on_mps_subterms} is flexible regarding the order of additions.

Another optimization can be achieved by reusing intermediate results. We first notice that one can write $G_{\mu \sigma, \nu \sigma^{\prime}}$ as:
\begin{equation}
G_{\mu \sigma, \nu \sigma^{\prime}} =  \zeta^{\mu \nu}G_{\mu \sigma}G_{\nu \sigma^{\prime}}
\end{equation}
where
\begin{equation}
G_{\nu \sigma^{\prime}} = \left(\sum^{L}_{r=1} \chi^{\nu}_r a^{\dagger}_{r,\sigma^{\prime}}\right) \left(\sum^{L}_{s=1}  \chi^{\nu}_s a_{s,\sigma^{\prime}}\right)
\end{equation}
and similarly for $G_{\mu \sigma}$. It indicates that for two sub-Hamiltonians $G_{a \tau, \nu \sigma^{\prime}}$ and $G_{b \kappa, \nu \sigma^{\prime}}$ that share the same latter two indices, the term $G_{\nu \sigma^{\prime}}$ can be factored out. Therefore, the intermediate state $G_{\nu \sigma^{\prime}} \ket{\psi}$, which is obtained from compressing the two elementary MPOs (layers) in $G_{\nu \sigma^{\prime}} \ket{\psi}$, can be reused. By applying this optimization, we reduce the computational cost by nearly half. Alg.~\ref{algo:mpo_on_mps} includes all these steps and illustrates the overall algorithm as pseudo-code.

In practice, we must take the kinetic term $T$ into account as well. The conventional MPO representation of $T$ has bond dimension $\mathcal{O}(L)$, which leads to an overall memory requirement of $\mathcal{O}(L^3 M^2)$ to store $T\ket{\psi}$ as MPS (without compression). We can improve on that situation using similar ideas as for the interaction term: We perform a spectral decomposition of $(t_{pq})$ in Eq.~\eqref{eq:ab_initio_H} and construct a sum of products of elementary MPOs with bond dimension $2$. Therefore, the memory requirement can be reduced to $\mathcal{O}(L M^2)$ for obtaining a compressed MPS. We explain these steps in detail in Appendix~\ref{app:kinetic}.

\begin{algorithm}[H]
\caption{Computing $V \ket{\psi}$ based on the THC-MPO}\label{algo:mpo_on_mps}
\begin{algorithmic}
\State \textbf{Input:} Initial state  $\ket{\psi}$ as MPS, all sub-Hamiltonians $G_{\mu \sigma, \nu \sigma^{\prime}}$ as MPOs
\State \textbf{Output:} $V \ket{\psi}$ as compressed MPS
\State $\ket{\phi} = 0$ 
\For{$\nu \in \{1,\dots,N\},\, \sigma' \in \{\uparrow,\downarrow\}$}
    \State Initialization: $\ket{\psi_{\nu \sigma^{\prime}}} = \ket{\psi}$
    \For {each elementary MPO in $G_{\nu \sigma^{\prime}}$}
        \State $\ket{\psi_{\nu \sigma^{\prime}}}$ =  $\text{elementary MPO} \cdot \ket{\psi_{\nu \sigma^{\prime}}}$
        \State Compress $\ket{\psi_{\nu \sigma^{\prime}}}$
    \EndFor

    \For{$\mu \in \{1,\dots,N\},\, \sigma \in \{\uparrow,\downarrow\}$}
        \State Initialization: $\ket{\psi_{\mu \sigma, \nu \sigma^{\prime}}} = \frac{1}{2} \zeta^{\mu \nu}\ket{\psi_{\nu \sigma^{\prime}}}$
        \For {each elementary MPO in $G_{\mu \sigma}$}
            \State $\ket{\psi_{\mu \sigma, \nu \sigma^{\prime}}}$ = $\text{elementary MPO} \cdot \ket{\psi_{\mu \sigma, \nu \sigma^{\prime}}}$
            \State Compress $\ket{\psi_{\mu \sigma, \nu \sigma^{\prime}}}$
        \EndFor
        \State {$\ket{\phi} = \ket{\phi} + \ket{\psi_{\mu \sigma, \nu \sigma^{\prime}}}$}
        \State Compress $\ket{\phi}$
    \EndFor
\EndFor
\State Return $\ket{\phi}$
\end{algorithmic}
\end{algorithm}

\section{Numerical results and resource estimation}
\label{sec:results}

\subsection{Ground- and low-lying states finding}
\label{subsec: exp_results}

To benchmark the MPS-based Lanczos algorithm using our THC-MPO, we apply our method to the water molecule \ce{H2O} and the hydrogen chain \ce{H10} using the STO-6G basis. The electronic integrals and FCI reference are calculated by PySCF~\cite{sun2018pyscf, sun2020recent}; the tensor network calculation is implemented with PyTeNet~\cite{pytenet}, in which Abelian quantum number conservation laws (electron number and spin) are enforced. We chose these relatively small systems because they allow for easier analysis of error sources and algorithmic behavior. But even so, we will see that the memory advantage has been fully verified. To fully explore our approach's computational complexity capabilities, we plan to switch to high-performance computers and utilize parallel computing to benchmark them for large systems in the future, as discussed in Sec.~\ref{sec:parallel}.

We first present the results of the water molecule using the STO-6G basis, which leads to $7$ spatial orbitals ($14$ spinor orbitals). In this case, we limit the maximum MPS bond dimension for the Krylov vectors to 30. The THC rank $N$ for \ce{H2O} is set to $28$, resulting in the Frobenius norm error $\norm{v - v^{\prime}} \approx 3 \times 10^{-11}$, where $v^{\prime}$ is the Coulomb term reconstructed by THC tensors according to Eq.~\eqref{eq:thc}.

\begin{figure}
\centering
\includegraphics[width=0.6\textwidth]{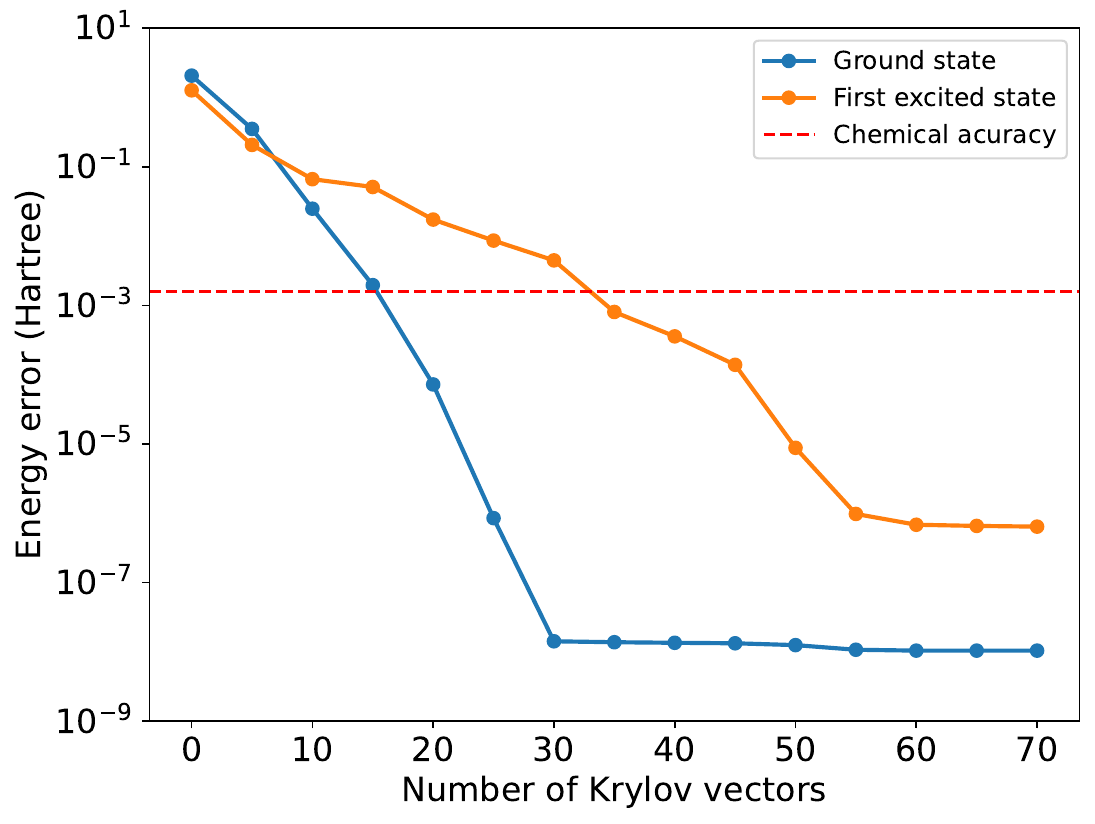}
\caption{Convergence of the water molecule's ground and first excited state calculation using the Lanczos algorithm based on our THC-MPO. We restart the iteration at the 45th step for the first excited state finding.}
\label{fig:water_energy}
\end{figure}

While the Lanczos algorithm performs well with a random initial state, selecting a proper initial state can significantly speed up convergence. In practice, we start from a state close to the target state, obtained from a heuristic guess or a low-cost algorithm. In this work, we simply use the Hartree-Fock state as the initial state for ground state finding, where paired electrons occupy the five lowest-energy molecular spatial orbitals. Additionally, we excite the highest occupied orbital in the Hartree-Fock state to serve as the initial state for finding the first excited state since the Hartree-Fock state is orthogonal to the ground state. As shown in Fig.~\ref{fig:water_energy}, we obtain the ground state and the first excited state within acceptable chemical accuracy (1.6 mHartree) using only $15$ and $35$ krylov vectors, respectively. The first excited state energy converges much slower because the gap between the first excited state and the second excited state is smaller than the one between the ground state and the first excited state. The energy error is obtained by comparison with the numerically exact value calculated by the FCI method in PySCF. 

However, a high-accurate THC decomposition as above is unnecessary because, first, the total error is also bounded by MPS truncation. Second, a high THC rank will increase the computational cost. A way to reduce computational cost at the expense of accuracy is using a smaller THC rank $N$. To explore this possibility and quantify the resulting error, we study the hydrogen chain of ten atoms \ce{H10} with distances of 1.4 Angstrom in the STO-6G basis, which leads to 10 spatial orbitals. Allowing a ground state energy error of $3 \times 10^{-6}$ Hartree per atom, its THC rank is as low as $27$.

\begin{figure}
\centering
\includegraphics[width=0.6\textwidth]{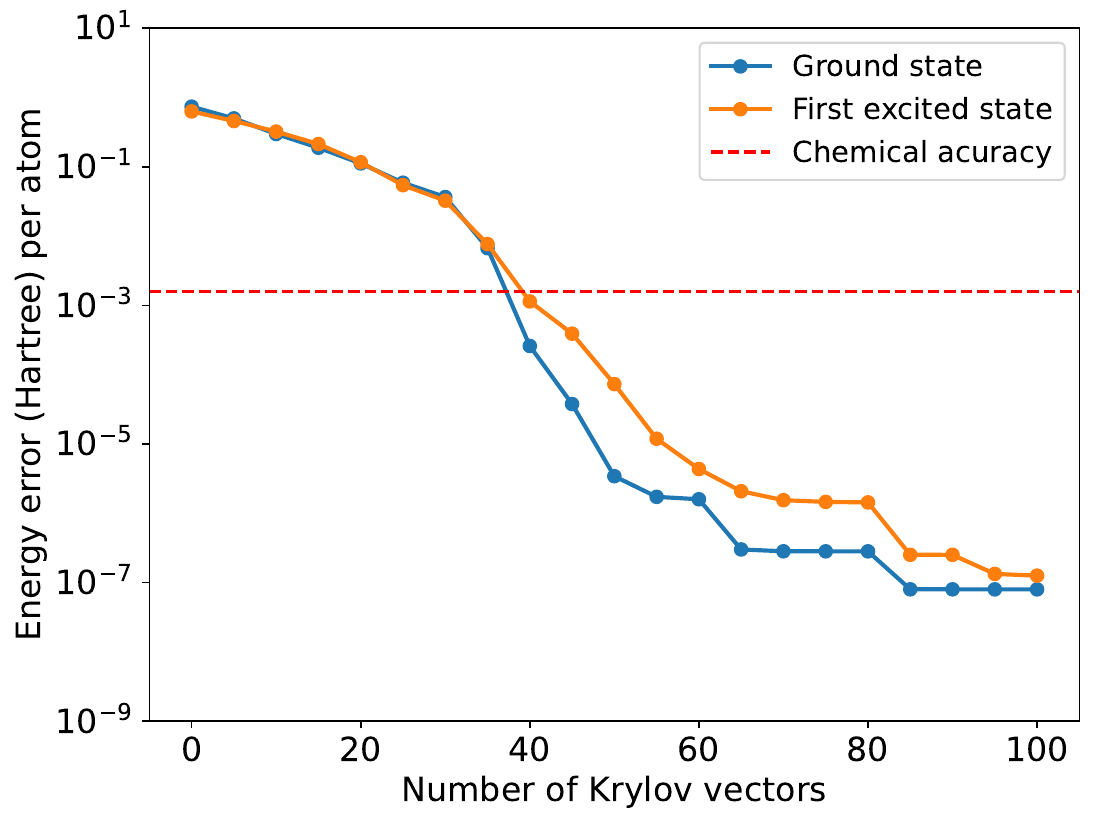}
\caption{Energy convergence for the hydrogen chain \ce{H10}. We restart the iteration at the 30th, 60th and 80th steps to improve the convergence. The Krylov space is obtained via THC-MPO, while the resulting energy of the approximated ground- and low-lying states is calculated according to the exact Hamiltonian.}
\label{fig:H10_energy}
\end{figure}

The MPS bond dimensions for representing Krylov vectors are capped at 250. Like the previous example, we again use the Hartree-Fock ground and single-excited states as the initial states for the Krylov method. Interestingly, when using the exact Hamiltonian to calculate the energy expectation value for the approximated ground state:
\begin{equation}
E_{\text{avg,Krylov}} = \bra{\psi_{\text{approx}}} H_{\text{exact}} \ket{\psi_{\text{approx}}},
\end{equation}
where $\ket{\psi_{\text{approx}}}$ is obtained by the THC-MPO-based Krylov method and $H_{\text{exact}}$ is the exact Hamiltonian,
the resulting energy error is smaller than the error introduced by the THC approximation. While the THC approximation leads to an energy error of around $3 \times 10^{-6}$ Hartree per atom, we can obtain the ground and first excited state with energy error $\sim 10^{-7}$ Hartree per atom, as illustrated in Fig.~\ref{fig:H10_energy}. This indicates that accurate results could still be obtained using the THC-MPO, even when choosing a smaller THC rank that introduces non-negligible errors. 

The results also suggest that although a large number of truncations is required to implement Eq.~\eqref{eq:H_on_mps_subterms}, the MPS truncations introduce only a minor error. Intuitively, assuming that each $G_{\mu \sigma, \nu \sigma^{\prime}} \ket{\psi}$ term admits a relative error $\epsilon$, the summation of them also admits a relative error $\epsilon$, especially when allowing larger bond dimensions during reduction (and compress the final bond dimension back to $M$). Despite errors introduced by compressing the terms $G_{\mu \sigma, \nu \sigma^{\prime}} \ket{\psi}$, additional errors also arise during the subsequent reduction (summation) process. Again, Assuming that pairwise addition-compression of these terms each results in a relative error $\epsilon'$, it follows that each hierarchical reduction level similarly contributes a relative error of $\epsilon'$. Given there are $\log(4N^2) = 2\log(N) + 2$ reduction levels (as shown in Fig.~\ref{fig:parallel}), the cumulative error can thus be bounded by $\epsilon' \sqrt{2\log(N) + 2}$,  assuming that the errors are all in different ''directions". Furthermore, since the reduction only manipulates the canonical MPS, increasing the bond dimension for this process will significantly enhance the accuracy without costing too much computational resources. Additionally, since the final MPS $\ket{\psi}$ closely approximates the ground state (or other low-lying eigenstates), the bond dimension required to accurately represent the resulting MPO-MPS is expected to remain moderate. Therefore, many sub-terms should not contribute a much larger error.

\subsection{Time evolution using global Krylov method}
\label{subsec: exp_evo_results}

We also study the Krylov subspace time evolution based on our THC-MPO, where we set the subspace dimension to $4$, leading to a single step error $\mathcal{O}(\delta^4)$ and total error $\mathcal{O}(\delta^3)$ for a fixed duration $T$ for a single time step size $\delta$. We apply the global Krylov method to the Ammonia molecule \ce{NH3} in the STO-6G basis, which leads to 8 spatial orbitals (16 spinor orbitals). The THC rank $N$ for \ce{NH3} is set to $36$, resulting in the Frobenius norm error $\norm{v - v^{\prime}} \approx 4\times 10^{-12}$. The initial state is defined as $\ket{\psi(t=0)} = a_{3,\uparrow} \ket{\psi_0}$, where a spin-up electron is annihilated from the third spatial orbital of the ground state. Three factors determine the accuracy: SVD cutoff (bond dimension limitation), time step size $\delta$, and the THC error from the THC factorization. The THC error is negligible for the \ce{NH3} molecule since the THC rank $N = 4.5L$ results in a very accurate approximation.


\begin{figure}
    \includegraphics[width=0.6\textwidth]{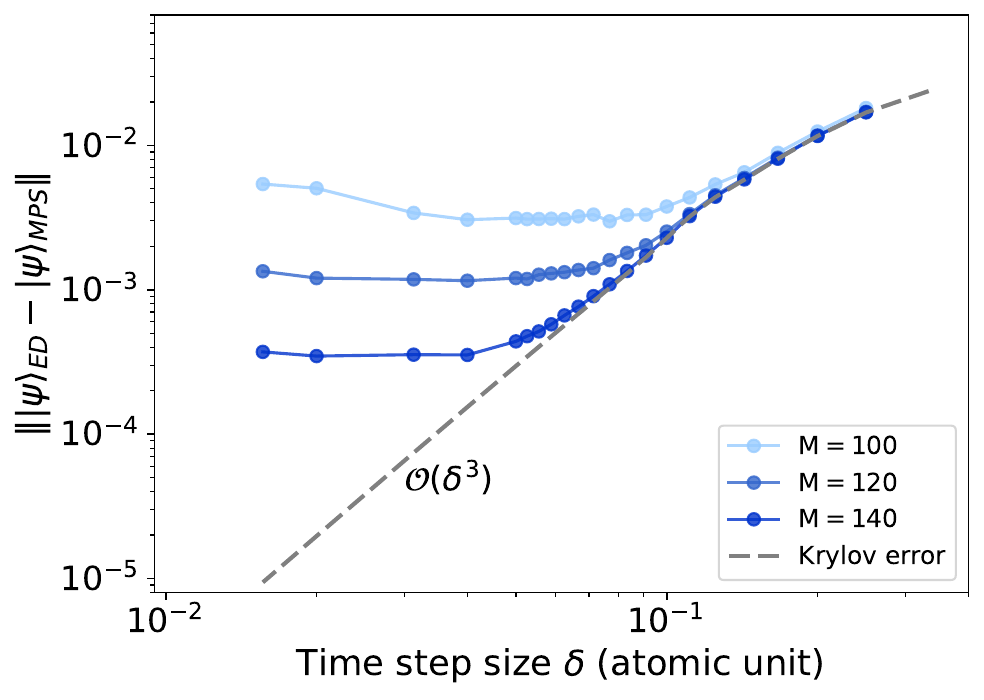}
    \caption{Time evolution errors of duration $T = 1$ atomic unit for \ce{NH3} when using the global Krylov method based on our THC-MPO for various bond dimensions $M$, plotted as functions of the time step sizes. The errors are measured by the distance $\norm{\ket{\psi}_{\text{ED}} - \ket{\psi}_{\text{MPS}}}$ between the states from our numerical method and the reference time-evolved quantum state obtained by ED. This metric is also used to measure the Krylov errors.}
    \label{fig:NH3_time}
\end{figure}

As depicted in Fig.~\ref{fig:NH3_time}, we measure the time evolution error for duration $T = 1$ atomic unit (a.u.) for different step sizes $\delta$ and maximum bond dimensions. The behavior of the errors can be explained well: As expected, the Krylov error dominates the overall error for larger time step sizes. Conversely, the $\mathcal{O}(\delta^3)$ scaling leads to small Krylov errors when the step size $\delta$ is reduced, causing the truncation error to dominate the overall error. To balance efficiency and accuracy, one can reach a sweet spot where the truncation error is comparable to the Krylov error. In Fig.~\ref{fig:NH3_time}, it occurs where the total error curve converges with the Krylov error curve. For the case $N=4$, one can observe that when setting $M = 140$, a step size of $\delta \in [0.05,\,0.1]$ a.u. appears to be optimal.

\begin{figure}
    \includegraphics[width=0.7\textwidth]{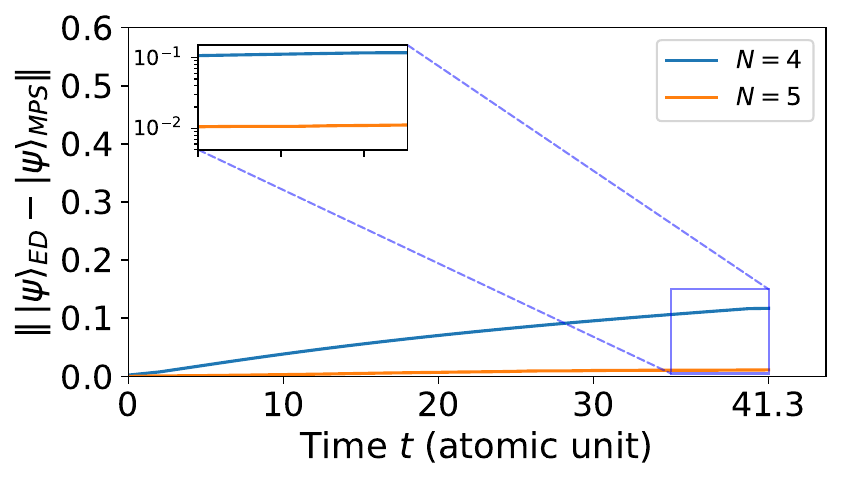}
    \caption{Time evolution errors of duration $T = 41.3$ atomic unit ($\approx$ 1 femtosecond) for \ce{NH3}, plotted as functions of the evolution time $t$.}
    \label{fig:NH3_longtime}
\end{figure}

It is also meaningful to enlarge the Krylov subspace size and examine whether it would enhance the accuracy as predicted. Specifically, we calculate the time evolution for duration $T = 41.3$ a.u. (1 femtosecond) with $M=140$, for both subspace size $N=4$ and $N=5$; the step size is set as $\delta = 0.1$. As illustrated in Fig.~\ref{fig:NH3_longtime}, the cumulative wavefunction error is $0.131$ for $N=4$ but only $0.011$ for $N=5$, indicating that the accuracy is enhanced by a factor of $10$ when adding one more vector. For this time step size $\delta = 0.1$, such a reduction is consistent with the expected error scaling $\mathcal{O}(\delta^N)$.

\subsection{Memory consumption comparison}
\label{subsec: memory}

\begin{figure}
\centering
\includegraphics[width=0.6\textwidth]{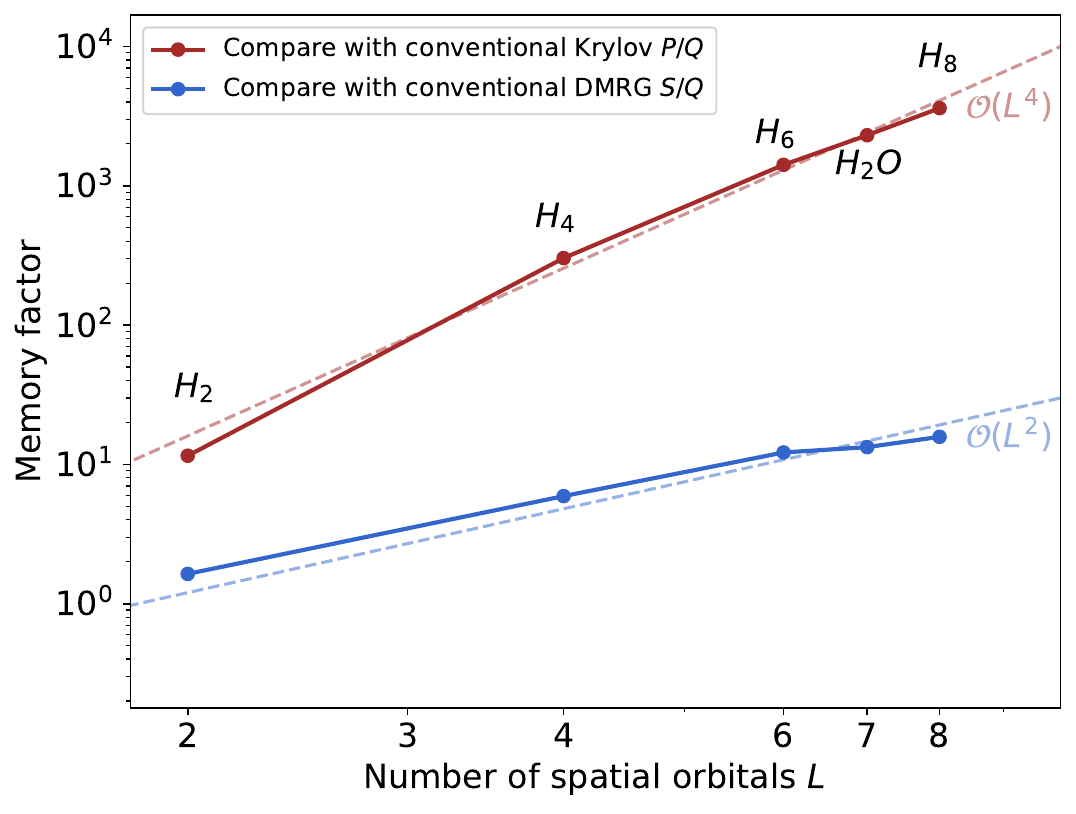}
\caption{Comparison of memory consumption for the Krylov method based on the conventional MPO versus the THC-MPO (red), as well as DMRG algorithm versus the Krylov method based on the THC-MPO (blue). The maximum bond dimensions are 4, 16, 60, 70, and 80, respectively. The dotted lines representing $\mathcal{O}(L^4)$ and $\mathcal{O}(L^2)$ demonstrate that the scaling of $P/Q$ and $S/Q$ aligns well with the theoretical prediction.}
\label{fig:memory}
\end{figure}

One of the advantages of our THC-MPO method is the significantly reduced memory cost by a factor $\mathcal{O}(L^4)$, We separately monitored memory consumption to store intermediate MPS in the Krylov algorithm based on the conventional MPO and the THC-MPO to test this prediction in our numerical experiments. We denote memory consumption when using conventional MPOs as $P$, and when using THC-MPOs as $Q$. Fig.~\ref{fig:memory} shows the quotient $P/Q$ for the water molecule and hydrogen chains. By studying systems of different sizes, one clearly observes the predicted $\mathcal{O}(L^4)$ scaling difference. For example, considering the hydrogen chain of eight atoms, the memory required for storing an intermediate state $H \ket{\psi}$ calculated with the conventional MPO amounts to 12586 MB. In contrast, only 3.49 MB is needed when using the THC-MPO method, leading to a factor $P/Q$ as large as 3606. We measure the memory cost by saving these intermediate MPS in HDF5 files and directly accessing their sizes. This case's maximum bond dimension is $80$, and we utilized double-precision complex numbers. Such a large memory usage is even too large to apply the Krylov methods on the \ce{H8} molecule. Therefore, in this respect, the \ce{H10} example has already proved the advantage over the original Krylov methods. 

The Krylov method based on THC-MPO also outperforms the DMRG algorithm in terms of memory usage. Theoretically, the DMRG algorithm requires $\mathcal{O}(L^3 M^2)$ memory (to store the left and right environment blocks), which is $\mathcal{O}(L^2)$ times larger than the THC-MPO-based Krylov method. As shown in Fig.~\ref{fig:memory}, we numerically compare $Q$ with the memory consumption $S$ for the DMRG algorithm, using the same MPS bond dimensions. The results suggest that our method requires significantly less memory than the DMRG algorithm, and the observed values agree with the theoretically predicted $\mathcal{O}(L^2)$ scaling. Since the TDVP method could be implemented within a framework similar to the DMRG algorithm, our method also outperforms TDVP in terms of memory consumption when simulating time evolution. We do not continue to increase the system size to measure more cases since the memory usage for conventional MPO methods rapidly exceeds our available memory (32 GB), and the results shown in Fig.~\ref{fig:memory} are sufficient to demonstrate the memory advantage of our method. Due to memory constraints, the runtime comparison for large systems is also infeasible.

\subsection{Computational complexity estimation}
\label{sec:estimation}

Besides memory consumption, the global Krylov methods based on our THC-MPO also perform better in terms of computational cost scaling than global Krylov methods using conventional MPOs. Here, we only present the summary; see Appendix~\ref{app:cost} for a detailed derivation.

\begin{figure}[h]
    \includegraphics[width=0.6\textwidth]{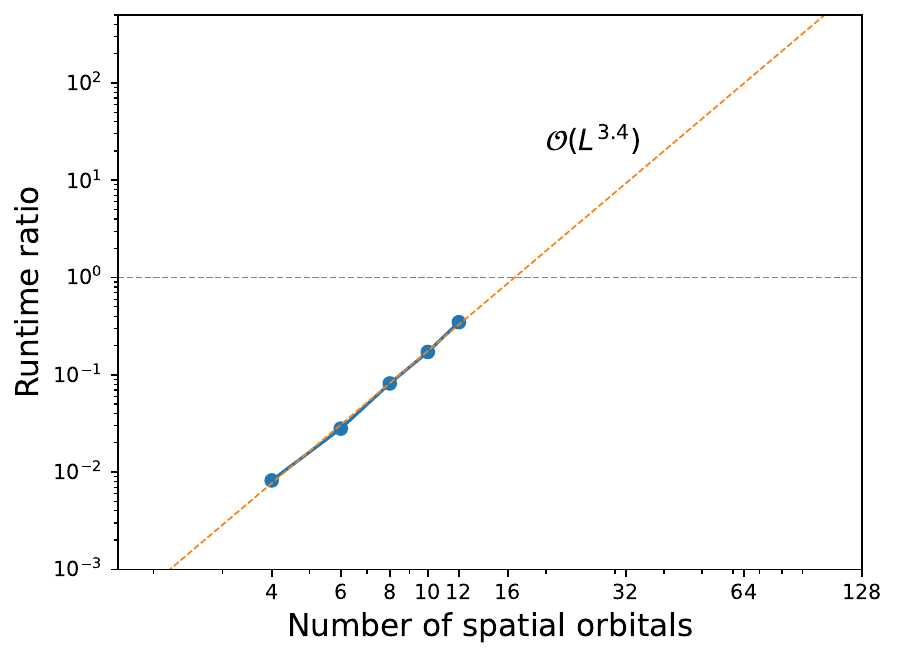}
    \caption{Log-log plot of the runtime ratio $t_{\text{THC}}/t_{o}$ as a function of the number of spatial orbitals.  
    The data points (blue markers) are fitted with a linear least-squares regression (orange dotted line). While the overall fitted slope is 3.4, the slope extracted from the last two data points rises to 3.8, indicating that the scaling trend converges toward our analytical prediction of $4$ as the system size increases.}
    \label{fig:runtime_compare}
\end{figure}

The primary contributor to the overall cost is obtaining compressed Krylov vectors. When using conventional MPO construction, renormalization is the most expensive step in compression. Since we have to handle the intermediate MPS with bond dimension $\mathcal{O}(L^2 M)$, the renormalization has an overall complexity of $\mathcal{O}(L^7 M^3)$ for all sites. In contrast, for global Krylov methods utilizing THC-MPO, we only need to deal with MPS with bond dimension $\mathcal{O}(M)$ since the bond dimension of each layer in the sub-terms $G_{\mu \sigma, \nu \sigma^{\prime}}$ is only $2$. Therefore, it costs $\mathcal{O}(L M^3)$ to obtain $G_{\mu \sigma, \nu \sigma^{\prime}} \ket{\psi}$ as compressed MPS, leading to an overall cost of $\mathcal{O}(L^3 M^3)$ for all $G_{\mu \sigma, \nu \sigma^{\prime}}\ket{\psi}$ (when assuming that the THC rank $N$ scales linearly with $L$). This computational cost has a large pre-factor; it could be around $10^5$ when taking hydrogen chains as an example. The large pre-factor leads to longer run times for small molecules; for example, on a 13th-generation Intel Core i7-1355U CPU (12 cores, 1.70 GHz base frequency), the average wall-time runtime is 56 seconds for computing a Krylov vector for our \ce{H2O} case, and 26 minutes is needed for a Krylov vector for our \ce{H10} case. The relatively slow performance can be attributed to our current implementation, which is not yet performance-oriented, and much of the multi-core capacity is left idle. Switching to a more efficient programming language and applying further optimizations should substantially improve the runtime. We are actively developing a more efficient implementation~\cite{chemtensor}.

Nevertheless, we expect an advantage for medium- and large-sized molecules due to their promising scaling gap $\mathcal{O}(L^4)$. To quantify the computational benefit of the THC-MPO method, we benchmark the wall-clock time required to compute the compressed MPS $H\ket{\psi}$. Denote the runtime with the explicit (original) Hamiltonian by $t_o$ and that with THC-MPO by $t_{\text{THC}}$, we report the runtime ratio $t_{\text{THC}} / t_o$, i.e., how many times faster the THC-MPO method is on hydrogen chains comprising 4 to 12 orbitals, as shown in Fig.~\ref{fig:runtime_compare}. Even though we only use a small bond dimension $M=30$,  the benchmark using conventional MPOs is still infeasible when $L \geq 14$ due to the large memory requirements. The curve admits a scaling $\mathcal{O}(L^{3.4})$, which is not as perfect accurate as our predicts, but the obtained runtime ratio can clearly reveal the ratio trend, that is, our THC-MPO method should be more efficient when the system size goes beyond $20$.

\subsection{A natural scalable parallelization scheme}
\label{sec:parallel}

Parallel computing has been effectively integrated into DMRG algorithms for quantum chemistry to take advantage of high-performance computing platforms. This integration has significantly enhanced the ability to study large molecular systems; various parallel schemes were proposed~\cite{zhai2021low, xiang2024distributed, DMRG-in-practice, large-scale-dmrg, wouters2014density, chan2016matrix, brabec2021massively}, and notable open-source packages like Block2 were developed~\cite{block2}. The Krylov method based on our THC-MPO can straightforwardly use the potential of parallel computing: to obtain $H\ket{\psi}$ following Eq.~\eqref{eq:H_on_mps_subterms}, each of the $4N^2$ sub-terms can be calculated and compressed independently, and the summation of these sub-terms can also be performed in parallel by a reduction. 

\begin{figure*}[hbt!]
\centering
\begin{tikzpicture}

    \draw[rounded corners=3pt, line width=0.8mm, draw=green!35!blue] 
        (0,0) rectangle (1.8,0.8);
    \node at (0.85,1.2) {\fontsize{9pt}{10pt}\selectfont{Core 1}};
    
    \node at (0.9,0.4) {\fontsize{10pt}{10pt}\selectfont\textbf{\({G_{1\uparrow, 1\uparrow}\ket{\psi}}\)}};
   
    \draw[rounded corners=3pt, line width=0.8mm, draw=green!35!blue] 
        (2.4,0) rectangle (4.2,0.8);
    \node at (3.25,1.2) {\fontsize{9pt}{10pt}\selectfont{Core 2}};
    \node at (3.3,0.4) {\fontsize{10pt}{10pt}\selectfont\textbf{\({G_{2\uparrow, 1\uparrow}\ket{\psi}}\)}};

    \draw[rounded corners=3pt, line width=0.8mm, draw=green!35!blue] 
        (4.9,0) rectangle (6.7,0.8);
    \node at (5.8,1.2) {\fontsize{9pt}{10pt}\selectfont{Core 3}};
    \node at (5.8,0.4) {\fontsize{10pt}{10pt}\selectfont\textbf{\({G_{3\uparrow, 1\uparrow}\ket{\psi}}\)}};

    \draw[rounded corners=3pt, line width=0.8mm, draw=green!35!blue] 
        (7.4,0) rectangle (9.2,0.8);
    \node at (8.3,1.2) {\fontsize{9pt}{10pt}\selectfont{Core 4}};
    \node at (8.3,0.4) {\fontsize{10pt}{10pt}\selectfont\textbf{\({G_{4\uparrow, 1\uparrow}\ket{\psi}}\)}};

    \draw[rounded corners=3pt, line width=0.8mm, draw=green!35!blue] 
        (9.9,0) rectangle (11.7,0.8);
    \node at (10.8,1.2) {\fontsize{9pt}{10pt}\selectfont{Core 5}};
    \node at (10.8,0.4) {\fontsize{10pt}{10pt}\selectfont\textbf{\({G_{5\uparrow, 1\uparrow}\ket{\psi}}\)}};
    
    \draw[line width=1.5pt] (13,0.5) circle (0.5pt);
    \draw[line width=1.5pt] (13.4,0.5) circle (0.5pt);
    \draw[line width=1.5pt] (13.8,0.5) circle (0.5pt);

    \draw[rounded corners=3pt, line width=0.8mm, draw=green!35!blue] 
        (14.8,0) rectangle (16.6,0.8); 
    \node at (15.7,1.2) {\fontsize{9pt}{10pt}\selectfont{Core $4N^2$}};
    \node at (15.7,0.4) {\fontsize{10pt}{10pt}\selectfont\textbf{\({G_{N\downarrow, N\downarrow}\ket{\psi}}\)}};

    \node at (1, -0.8) {\fontsize{10pt}{10pt}\selectfont\text{\({\oplus}\)}};
    \node at (5.8, -0.8) {\fontsize{10pt}{10pt}\selectfont\text{\({\oplus}\)}};
    \node at (10.6, -0.8) {\fontsize{10pt}{10pt}\selectfont\text{\({\oplus}\)}};
    
    \Edge[Direct](1,-0.1)(1,-0.8 * 0.8)
    \Edge[Direct](5.8,-0.1)(5.8,-0.8 * 0.8)
    \Edge[Direct](10.6,-0.1)(10.6,-0.8 * 0.8)
    \Edge[Direct](15.7,-0.2)(14,-0.8 * 0.8)

    \node at (1, -2.5 * 0.8) {\fontsize{10pt}{10pt}\selectfont\text{\({\oplus}\)}};

    \Edge[Direct](1,-1.3 * 0.8)(1,-2.25 * 0.8)
    \Edge[Direct](1,-2.75 * 0.8)(1,-3.55 * 0.8)
    \node at (1, -3.8 * 0.8) {\fontsize{10pt}{10pt}\selectfont\text{\({\vdots}\)}};
    \Edge[Direct](1,-4.35 * 0.8)(1,-5.15 * 0.8)
    \node at (1, -5.5 * 0.8) {\fontsize{10pt}{10pt}\selectfont\text{\({V \ket{\psi}}\)}};

    \node at (10.6, -2.5 * 0.8) {\fontsize{10pt}{10pt}\selectfont\text{\({\vdots}\)}};

    \Edge[Direct](10.6,-1.3 * 0.8)(10.6,-2.25 * 0.8)

    \Edge[Direct](5.4,-1.1 * 0.8)(1.6,-2.3 * 0.8)

    \Edge[Direct](3.5,-0.1)(1.3,-0.8 * 0.8)
    \Edge[Direct](8.3,-0.1)(6.1,-0.8 * 0.8)
    \Edge[Direct](12.8,-0.2)(10.9,-0.8 * 0.8)

\end{tikzpicture}
\caption{Parallelization scheme for applying the Coulomb operator $V$ to a state $\ket{\psi}$ in MPS form according to Eq.~\eqref{eq:H_on_mps_subterms}. Each core is first assigned the sub-task to compute and compress an intermediate state $G_{\mu \sigma, \nu \sigma^{\prime}} \ket{\psi}$ as MPS. These are then aggregated through a reduction process. For simplicity, we assume that the high-performance computer is able to perform at least $4N^2$ cores; otherwise, a single core would handle several of the $G_{\mu \sigma, \nu \sigma^{\prime}} \ket{\psi}$ states.}
\label{fig:parallel}
\end{figure*}

More specifically, we propose a parallelism scheme as illustrated in Fig.~\ref{fig:parallel}. For each core, we first assign the task of computing and compressing one (or several) sub-terms $G_{\mu \sigma, \nu \sigma^{\prime}} \ket{\psi}$. The power of multiple cores can be perfectly utilized for this part. After this step, we add and compress these terms pairwise in parallel. It appears that some computational resources are idling during such a process, but the compression can utilize multiple cores for parallel computation when using packages like multithreaded LAPACK implementations~\cite{blackford2002updated}. Because the SVD and QR decomposition can be significantly sped up by parallel computing~\cite{demmel2012communication, jessup1994parallel}, the reduction part can utilize the power of parallel computing as well. Also, since each compressed term $G_{\mu \sigma, \nu \sigma^{\prime}} \ket{\psi}$ has already been canonical form (up to a factor), the MPS addition-compression process doesn't contain the QR decomposition, which makes the reduction inexpensive. Another bottleneck of parallel computing is communication \cite{zhai2021low}; an extra advantage of our parallel scheme is that communication only occurs during the reduction.

\begin{figure}
\centering
\includegraphics[width=0.6\textwidth]{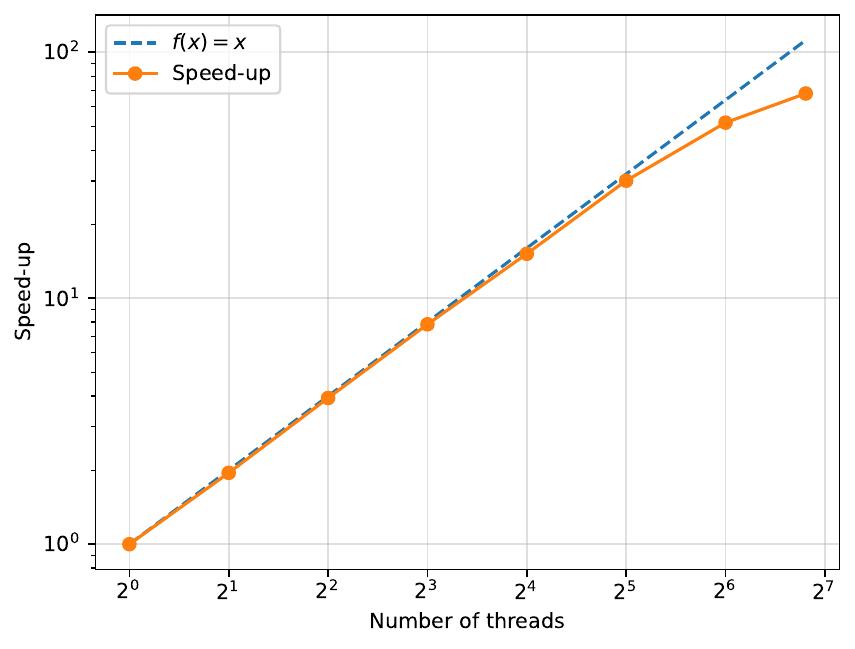}
\caption{The speedup is calculated by comparing the runtime using $K$ threads with that of a single thread. Although slightly imperfect due to factors such as cache contention, the results indicate that our parallel scheme efficiently leverages available computational resources.}
\label{fig:speed_up}
\end{figure}

As a preliminary demonstration (with more advanced systems and larger molecules planned for future work), we benchmark our parallel scheme on a 112-core node using OpenMP~\cite{dagum1998openmp}. Each thread (core) is tasked with calculating and compressing $G_{\mu \sigma, \nu \sigma^{\prime}} \ket{\psi}$. As displayed in Fig.~\ref{fig:speed_up}, we compare the runtime for computing $H\ket{\psi}$ with a single thread versus $K$ threads. The results indicate near-ideal speedups when multiple threads are utilized. Extending this approach to multiple nodes should also yield near-linear scaling because each node can achieve this speedup independently, and communication among nodes is only required once all nodes have completed their tasks.

Due to such an efficient and scalable parallel scheme, the parallel runtime scales as $\mathcal{O}(L M^3)$ under ideal parallelization conditions with $4 N^2$ available cores efficiently. Currently, advanced tensor network methods in quantum chemistry (e.g., DMRG) can utilize thousands of cores efficiently~\cite{brabec2021massively, zhai2021low}, the Krylov methods (Lanczos algorithm and Krylov time evolution method) based on THC-MPO have the potential to leverage cores scaling as $4 N^2$ with high efficiency, making it possible to surpass the current state of the art in CPU utilization. Note that each sub-task shown in Fig.~\ref{fig:parallel} can also be implemented by multiple cores (e.g., a node), thereby further increasing the number of cores we can efficiently utilize and decreasing the reduction depth.

\section{Conclusions}
\label{sec:conclusions}
The THC-MPO approach allows us to implement Krylov subspace methods, e.g., the Lanczos algorithm and the global Krylov method for time evolution, with reduced memory usage and lower computational cost scaling. When compared to the Krylov method based on the conventional MPO representation, the memory advantage of THC-MPO is apparent, even for the smallest molecules. Moreover, it outperforms popular methods like DMRG and TDVP in terms of memory consumption, suggesting that THC-MPO can potentially enable simulations of even larger systems than currently reachable by DMRG or TDVP. While the benefit of computational cost is not immediate for small molecules due to large prefactors, we expect that the improvement will become significant for moderate and large molecular systems. We emphasize that the THC-MPO is essentially a proper decomposition. A group of MPOs with a small bond dimension can also be achieved by building an MPO for each term in the molecular Hamiltonian, but the final computational complexity will be as large as $\mathcal{O}(L^7)$ if we do so.

A cornerstone of our work is the compressed THC representation of the two-body integral tensor $v$. A promising research direction (complementary to the present study) could be the exploitation of sparsity structures of $v$, for example, due to localized orbitals or wavelet-type orbitals supported on a fine grid. Also, the tensor $v_{pqrs}$ (which originates from overlap integrals) is symmetric with respect to interchanges of $p \leftrightarrow q$, $r \leftrightarrow s$, and $(p, q) \leftrightarrow (r, s)$. The symmetries are passed on to the THC representation. It is worth exploring how to exploit such symmetries in our approach. We also noticed that our THC-MPO could help enable the computation of spectral functions~\cite{yang2020probing, jiang2021chebyshev, holzner2011chebyshev} for large-size molecular Hamiltonians when combining with the Chebyshev expansions, where multiplication-compression $H\ket{\psi}$ also remains a major bottleneck. With our THC-MPO method and optimized parallel computing implementations, our primary plan is to explore these ideas and the reachable system sizes in future works.

\begin{acknowledgement}
We thank Haibo Ma, Zhaoxuan Xie, Ke Liao, and Philipp Seitz for their helpful discussion. C.~M.\ acknowledges funding by the Munich Quantum Valley initiative, supported by the Bavarian state government with funds from the Hightech Agenda Bayern Plus.
\end{acknowledgement}

\appendix

\section{Error estimation for Krylov subspace time evolution}
\label{app:time_evo_error}
As discussed by Hochbruck and Lubich~\cite{kryloverror}, the error $\varepsilon_N$ incurred by the $N$th-order Krylov approximation of the time-evolution operator $e^{-itH}$ is bounded by:

\begin{equation}
\varepsilon_{N} \le 
\begin{cases}
10 \, e^{-4N^2 /(5 W\delta)}, 
& \text{if } \sqrt{W\delta} \le N \le \frac{W\delta}{2}, \\[0.5em]
10 \left(\frac{W\delta}{4}\right)^{-1} e^{-\frac{W\delta}{4}} \left(\frac{eW\delta}{4N}\right)^{N}, 
& \text{if } N \ge \frac{W\delta}{2},
\end{cases}
\label{eq:error_bond}
\end{equation} where $W$ is the spectral width, and $\delta$ is the time step size. Even though the original formula is derived for a Hermitian negative semi-definite matrix $A$ with eigenvalues in the range $\left[-W,\,0\right]$,  one may equivalently simulate the shifted operator $A + \frac{W}{2} I$, which only results in a global phase $e^{-i \frac{W\delta}{2}}$.

We can estimate the theoretical error upper bounds for different Krylov subspace sizes according to Eq.~\eqref{eq:error_bond}. A satisfactory accuracy (error $< 10^{-4}$) for a single-step evolution can be obtained for $N=4$ with step size $W\delta < 0.2$, or for $N=5$ when $W\delta < 0.5$. Higher precision can be achieved either by enlarging the subspace or by reducing the time step size. For instance, the accuracy level $10^{-7}$ can be achieved by time step size $W\delta = 0.1$ when $N=5$, or $ W\delta = 1$ when $N=8$. Note that even though the Krylov vector could reach very high accuracy, the algorithm could suffer from the MPS truncation error. The most efficient way is to simulate the time evolution using a time step size or subspace size where the error incurred by the Krylov vector is comparable to the MPS truncation error.

\section{Alternative MPS compression methods}
\label{app:alternative}

The density matrix method~\cite{ma2024approximate} is a newly proposed error-controlled method. Its computational complexity is $\mathcal{O}(L^5 M^3)$ for a conventional MPO, and it can be reduced to $\mathcal{O}(L^3 M^3)$ when employing our THC-MPO method. Its memory cost is $\mathcal{O}(L^4 M^2)$ to store the $L_i$ matrices, and it can be reduced to $\mathcal{O}(L M^2)$ by our THC-MPO method.

The zip-up method~\cite{Stoudenmire_2010} trades computational efficiency for controllability of the approximation error. The reason is that the zip-up method uses a non-orthogonal basis~\cite{ma2024approximate}. Therefore, its cost is reduced, but the magnitude of the error is not explicitly known. Applied to the molecular Hamiltonian, its computational complexity scales as $\mathcal{O}(L^5 M^2 + L^3 M^3)$, which can be reduced to $\mathcal{O}(L^3 M^3)$ when employing our THC-MPO method. Its memory cost is $\mathcal{O}(L^2 M^2)$, reducible to $\mathcal{O}(M^2)$ by our THC-MPO method. Furthermore, the memory required to store the conventional MPO can grow prohibitively large due to the memory cost scaling $\mathcal{O}(L^4)$, especially for a large number of spatial orbitals (e.g., larger than 100). By contrast, storing a single THC-MPO term $G_{\mu \sigma, \nu \sigma^{\prime}} \ket{\psi}$ demands only $\mathcal{O}(L)$ memory.

The variational method~\cite{schollwock2011density} is prone to getting trapped in local minima, and hence, a proper initial guess is needed. One may use an inaccurate compression, such as the zip-up method, for this initial guess, and then perform a further variational optimization, as suggested in~\cite{PAECKEL2019167998}. In this scenario, the dominant cost will not stem from the variational method itself.

We would like to emphasize that our THC-MPO method does not engage in a competition between the SVD method and others, but provides potential improvements, including memory cost, computational complexity, and parallelizability, for all of the MPS compression methods. As more innovative MPS compression methods will be proposed in the future, we can always check whether our THC-MPO method can be combined with them.

\section{Local annihilation and creation operators}
\label{app:local_op}
Two spinor orbitals are contained in each spatial orbital; therefore, the Pauli-$Z$ operator is also needed inside each site when employing the Jordan-Wigner transformation. Ordering the spin-up before the spin-down, the local annihilation operators for site $s$ are written as:
\begin{equation}
\begin{aligned}
b_{s, \uparrow} &= c \otimes I_2, \\
b_{s, \downarrow} &= Z \otimes c,
\end{aligned}
\end{equation}
where $c$ is defined as the $2 \times 2$ matrix $c \equiv \left(\begin{smallmatrix}
    0 & 1 \\
    0 & 0
\end{smallmatrix}\right)$, and $I_2$ denotes the identity matrix of size $2 \times 2$. The creation operators can be readily obtained by taking the conjugate transpose of the annihilation operators.

\section{Decomposition of the kinetic term}
\label{app:kinetic}

The matrix $(t_{pq})$ of one-body integrals is real symmetric and can thus be diagonalized via an orthogonal matrix $(u_{pi})$ of eigenvectors and corresponding eigenvalues $\lambda_i$:
\begin{equation}
\label{eq:t_diagonalization}
t_{pq} = \sum^{L}_{i=1} u_{pi} \lambda_i u_{qi} \quad \text{for all } p, q = 1, \dots, L.
\end{equation}
Inserted into the kinetic term in Eq.~\eqref{eq:ab_initio_H}, we directly obtain:
\begin{equation}
T = \sum^{L}_{i=1} \sum_{\sigma \in \{\uparrow, \downarrow\}} \underbrace{\lambda_i \left(\sum^{L}_{p=1} u_{pi} a^{\dagger}_{p, \sigma}\right) \left(\sum^{L}_{q=1} u_{qi} a_{q, \sigma}\right)}_{= T_{i, \sigma}}.
\end{equation}
For each sub-term $T_{i, \sigma}$, one can construct elementary MPOs for $\sum^{L}_{p=1} u_{pi} a^{\dagger}_{p, \sigma}$ and $\sum^{L}_{q=1} u_{qi} a_{q, \sigma}$ in the same way as in Eq.~\eqref{eq:elementary_MPO}. Therefore, $T_{i, \sigma}$ is a product of two MPOs with individual bond dimensions 2. Since the kinetic term is the sum of the sub-terms $T_{i, \sigma}$, the operation $T\ket{\psi}$ can be performed analogously to the Coulomb interaction by sequential summation and compression of the states $T_{i, \sigma} \ket{\psi}$ in MPS form. In total, there are $2L$ sub-terms for the kinetic part, which is relatively small compared to the $\mathcal{O}(L^2)$ sub-terms arising from the Coulomb interaction. Moreover, note that the spectral decomposition \eqref{eq:t_diagonalization} is numerically exact, while the THC representation of the Coulomb overlap integrals in Eq.~\eqref{eq:thc0} is an approximation in general.

\section{Computational cost estimate for computing Krylov vectors}
\label{app:cost}

There are three primary steps to compute $H\ket{\psi}$ in compressed MPS form: multiplying $H$ with $\ket{\psi}$, renormalization, and truncation by SVD. Regarding the conventional MPO-based method, the most expensive step is renormalization, which contains QR decompositions and the subsequent absorption of the $R$ matrices from the QR decomposition into the next site. The QR decomposition on tensors of shape $(2, L^2M, L^2M)$ leads to a cost of $\sim \frac{10}{3}L^6 M^3$ floating point operations if the Householder reflection method is utilized~\cite{golub2013matrix}. Such a decomposition results in an $R$ matrix of shape $(L^2M, L^2M)$, absorbing it into the next site costs $\sim 2L^6M^3$. The leading term in computational cost is $\sim \frac{32}{3}L^7M^3$ floating point operations for all $2L$ sites (spin-orbitals). The SVD cost is minor: starting from the very left or right side, one of the two MPS virtual bonds has already been reduced to $M$, leading to a tensor of shape $(2, M, L^2 M)$. Applying an SVD of such a tensor only costs $\mathcal{O}(L^2 M^3)$, which is much smaller than the cost from the QR decomposition. The asymptotic scaling $\mathcal{O}(L^7M^3)$ is a significant hurdle when applying global methods to large systems. Typically, the maximum MPO bond dimensions exceed $L^2$, so we provide only a rough estimation to offer some intuition. 

For THC-MPO, we first discuss the computational complexity of evaluating a sub-term $G_{\mu \sigma, \nu \sigma^{\prime}}\ket{\psi}$, for which we execute multiplication and compression layer by layer as discussed in~\ref{subsec:thc-krylov}. After multiplying a layer with the current MPS, the shape of the resulting temporary MPS tensors is $(2, 2M, 2M)$ since the MPO bond dimension for each layer is 2. Thus, employing a single-site QR decomposition costs $\sim \frac{80}{3}M^3$ floating point operations. Subsequently, absorbing the $R$ matrix of shape $(2M, 2M)$ into the next site costs $\sim 16M^3$. Finally, we apply an SVD to truncate the intermediate MPS. Similarly to the conventional case, one of the two MPS virtual bonds has already been reduced to $M$, leading to a tensor of shape $(2, M, 2M)$. Applying an SVD of such a tensor costs $\sim 56M^3$~\cite{golub2013matrix} using the divide-and-conquer method implemented in LAPACK~\cite{gu1996efficient, anderson1999lapack}, and absorbing the obtained matrix into the next site costs $\sim 8M^3$. In summary, for each of the four layers in $G_{\mu \sigma, \nu \sigma^{\prime}}$ it costs $\sim 214LM^3$ to compress the intermediate MPS for all $2L$ sites, leading to $\sim 10^3LM^3$ floating point operations to obtain $G_{\mu \sigma, \nu \sigma^{\prime}} \ket{\psi}$ in compressed MPS form. To implement Eq.~\eqref{eq:H_on_mps_subterms}, one needs to execute $4 N^2$ times multiplication-compression where $N$ is the THC rank which scales linearly with system size. Taking the hydrogen chain with $N = 3 L -3$ as an example, this leads to a total cost of $\sim 10^4 L^3 M^3$ floating point operations, considering the optimization in which we re-use the later half of $G_{\mu \sigma, \nu \sigma^{\prime}}$, as mentioned in Sec.~\ref{subsec:thc-krylov}. We also need to implement $4 N^2 -1 $ times addition-compression, but its cost is negligible in comparison since QR decomposition is not necessary.

Even though the complexity estimation here is approximate since we treat all bond dimensions as $M$ for simplicity and might need larger $M$ for desired accuracy, the asymptotic scaling gap $\mathcal{O}(L^4)$ is faithfully captured. Comparing the cost $\sim 10 L^7 M^3$ for a conventional MPO with $ \sim 10^4 L^3 M^3$ for the THC-MPO method, the crossover point is estimated to occur when $L$ is in the range of a few tens.

\bibliography{references}

\end{document}